\documentclass[aps,pre,amsmath,amssymb,amsfonts,lengthcheck,twocolumn,superscriptaddress]{revtex4-2}
\usepackage{graphicx}
\usepackage{subfigure}
\usepackage{amsthm}
\usepackage{verbatim}
\usepackage{dcolumn}
\usepackage{bm}
\usepackage{epsf}
\usepackage{color}
\usepackage[colorlinks=true,citecolor=blue,linkcolor=blue,urlcolor=blue]{hyperref}%
\usepackage{xcolor}
\usepackage{dsfont}
\usepackage{tikz}
\usepackage{todonotes}
\usepackage{multirow}
\usepackage{xfrac}
\usepackage{makecell}

\newcommand{\bra}[1]{\left\langle #1\right|}
\newcommand{\ket}[1]{\left|#1\right\rangle}
\newcommand{\braket}[2]{\left\langle #1|#2\right\rangle}

\newcommand{\tr}[1]{\mathrm{tr}\left\{#1\right\}}

\newcommand{\pd}{\partial}

\newcommand{\e}[1]{\exp{\left(#1\right)}}

\newcommand{\id}{\mathbb{I}}

\newcommand{\bla}{bla\\bla\\bla\\bla\\bla}

\newcommand{\mc}[1]{\mathcal{#1}}

\newcommand{\mrm}[1]{\mathrm{#1}}

\DeclareMathAlphabet\mathbfcal{OMS}{cmsy}{b}{n}

\makeatletter
\newcommand{\currentfontsize}{\f@size pt}
\makeatother

\makeatletter
\newcommand\footnoteref[1]{\protected@xdef\@thefnmark{\ref{#1}}\@footnotemark}
\makeatother

\begin{document}

\title{Quantum thermodynamics of Gross-Pitaevskii qubits}

\author{Sebastian Deffner}
\email{deffner@umbc.edu}
\affiliation{Department of Physics, University of Maryland, Baltimore County, Baltimore, MD 21250, USA}
\affiliation{Quantum Science Institute, University of Maryland, Baltimore County, Baltimore, MD 21250, USA}
\affiliation{National Quantum Laboratory, College Park, MD 20740, USA}

\begin{abstract}
What are the resources that can be leveraged for a thermodynamic device to exhibit genuine quantum advantage? Typically, the answer to this question is sought in quantum correlations. In the present work, we show that quantum Otto engines that operate with nonlinear qubits significantly outperform linear engines. To this end, we develop a comprehensive thermodynamic description of nonlinear qubits starting with identifying the proper thermodynamic equilibrium state. We then show that for ideal cycles as well as at maximum power the efficiency of the nonlinear engine is significantly higher. Interestingly, nonlinear dynamics can be thought of as an effective description of a correlated, complex quantum many body system. Hence, our findings corroborate common wisdom, while at the same time propose a new design of more efficient quantum engines.
\end{abstract}

\maketitle

\section{Introduction}

A rather well-known quote credited to Stanilaw Ulam reads \cite{Campbell1985CACM},
\begin{quote}
\emph{Using a term like nonlinear science is like referring to the bulk of zoology as the study of non-elephant animals.}
\end{quote}
In essence, this statement expresses the fact that while we often use linear systems to describe the universe, most phenomena in the real world are nonlinear. 

A notable exception to this rule of nature are quantum mechanical systems, for which it can be shown that the dynamics must fundamentally be linear \cite{Jordan2009JPCF}. This fact has been extensively verified in experiments, see for instance Refs.~\cite{Walsworth1990PRL,Majumder1990PRL,Forstner2020Optica}. However, in  complex quantum many-body systems, the properties of collective excitations can often be effectively described by non-linear Schr\"odinger equations \cite{DalFavero2024PRA}. Among these, the Gross-Pitaevskii equation \cite{Gross1961,Pitaevskii1961} is probably most widely known. In space representation, it can be written as,
\begin{equation}
\label{eq:GPE}
i\hbar\, \dot{\psi}(x,t)=H(x,t) \psi(x,t) +\kappa \left|\psi(x,t)\right|^2\psi(x,t)
\end{equation}
where $H(x,t)$ is the Hamiltonian, and $\kappa$ measures the ``strength'' of the cooperative quantum effects. In Bose-Einstein condensates, $\kappa$ is essentially the scattering length \cite{Gross1961,Pitaevskii1961}, with $\kappa<0$ for repulsive and $\kappa>0$ for attractive interactions. However, Eq.~\eqref{eq:GPE} does not only describe cold quantum gases, but has also found applications in, e.g., nonlinear optics \cite{Rand2010} and in plasma physics \cite{Ruderman2002}. Interestingly, Eq.~\eqref{eq:GPE} has also been generalized to nonlinear Dirac equations \cite{Soler1970PRD,Cooper2010PRE,Mertens2012PRE}, which describe self-interacting fermions.

More generally, nonlinear wave equations like Eq~\eqref{eq:GPE} are obtained as meanfield solutions of interacting field theories. For instance, when including more than 2-body interactions one obtains the 
Kolomeisky equation \cite{Kolomeisky2000PRL} with a qunitic nonlinearity. In the limit of Bose liquids, the nonlinear terms even becomes logarithmic in the wave function  \cite{Meyer2014PRA}.

From the point of view of quantum information processing, nonlinear variants of the Schr\"odinger equation are very appealing as such dynamics can facilitate processes that are not possible in linear systems, such as, e.g., perfectly distinguishing non-orthogonal states \cite{Childs2016PRA} or synchronization \cite{Zhang2020MPLB,Shen2023PRA}. Thus, in recent years, significant work has been dedicated to the development of ``nonlinear quantum computation'' \cite{Meyer2014PRA,Lacy2018QIP,Chiew2019QIP,Holmes2023PRA,Geller2023CTP,Geller2023AQT,Deiml2024arXiv}. The effectively nonlinear dynamics are then generated by, for instance, coupling Bose-Einstein condensates \cite{Byrnes2015,Xu2022PRR,Geller2024AQT,Grossardt2024arXiv}, building qubits from chaotic orbits \cite{Geller2023SR}, intermittent measurements \cite{Kalman2018PRA,Sakaguchi2023NC}, or nonlinear optics \cite{Yang2008CTP,Chang2014,Gu2016AOP,Scala2024CP}. On the more theoretical side, nonlinear dynamics have been explored in the context of quantum speed limits \cite{Campo2021PRL,Deffner2022EPL}, quantum metrology \cite{Beau2017PRL,Deffner2025QST}, and shortcuts to adiabaticity \cite{Deffner2014PRX,Chen2016PRA,Zhu2020PRA,Zhu2021PRA,Zhu2023PRA}.

In the present analysis, we continue the development of a more comprehensive thermodynamic description of nonlinear qubits that we started in Ref.~\cite{Deffner2025EPL}. To this end, we consider the most general qubits that are subject to arbitrary nonlinear dynamics. We consider such qubits as a thermodynamic system and determine the quantum state of the corresponding canonical ensemble. Note that we have seen in previous analyses \cite{Deffner2025QST} that the Gibbs state is generally not stationary under the nonlinear dynamics. Hence, the canonical ensemble is not simply given by a Gibbs state, but rather we have to determine the actual quantum state representing thermodynamic equilibrium. 

We find that the nonlinear interaction leads to an increase in internal energy and entropy, but decrease in maximal heat capacity (as a function of temperature). We then solve for the efficiency of ideal as well as endorversible Otto cycles, and we find that, for infinitely slow as well as at maximum power in the endoreversible regime, the nonlinear qubit engine outperforms the linear version. Thus, as a main result, we find that nonlinear qubits possess additional resources that can be leveraged in thermodynamic performance.

\section{General nonlinear qubits}

We start by establishing notions and notations for general qubits undergoing nonlinear dynamics. In the following, we consider quantum dynamics described by the nonlinear Schr\"odinger equation \cite{Childs2016PRA}
\begin{equation}
\label{eq:nonlin_sch}
i \, \pd_t \ket{\psi_t}=H_t\,\ket{\psi_t}+ K \ket{\psi_t}
\end{equation}
where $H_t$ is the usual, Hermitian, time-dependent Hamiltonian. To avoid clutter in the formulas, we will work in units for which $\hbar=1$. Further, $K$ describes a general nonlinearity of the form
\begin{equation}
\bra{x} \left(K \ket{\psi_t}\right)= \mc{K}(|\braket{x}{\psi_t}|)\,\braket{x}{\psi_t}\,.
\end{equation}
Almost all nonlinear quantum dynamics of relevance can be written in this form. For instance, for the Gross-Pitaeveskii equation \cite{Gross1961,Pitaevskii1961} we simply have $\mc{K}(|\braket{x}{\psi_t}|)=\kappa\,|\braket{x}{\psi_t}|^2$, and for the Kolomeisky equation \cite{Kolomeisky2000PRL}, $\mc{K}(|\braket{x}{\psi_t}|)=\kappa\,|\braket{x}{\psi_t}|^4$.

\paragraph*{Hamiltonian and nonlinear dynamics}

We now consider the most general qubit Hamiltonian
\begin{equation}
\label{eq:H_qubit}
H=\vec{B}\cdot \vec{\sigma}
\end{equation}
where $\vec{B}=(\xi,\chi,\zeta)$ is the magnetic field, and $\vec{\sigma}=(\sigma_x,\sigma_y,\sigma_z)$ is the Pauli vector. 

In the following, it will be convenient to write the quantum state $\ket{\psi_t}$ in its Bloch representation,
\begin{equation}
\label{eq:bloch}
    \rho_t=\ket{\psi_t}\bra{\psi_t}=\frac{1}{2}\left(\id_2+\vec{r}_t\cdot\vec{\sigma}\right)\,,
\end{equation}
where $\vec{r}_t=(x_t,y_t,z_t)$ is the Bloch vector. Note that for pure states $|\vec{r}|=1$. It is worth emphasizing that the nonlinear Schr\"odinger equation \eqref{eq:nonlin_sch} is purity preserving, i.e., if $|\vec{r}_0|=1$ then $|\vec{r}_t|=1$ for all $t$. Note that the corresponding propagator is \emph{not} unitary, since the dynamics is \emph{not} linear.

In the Bloch representation \eqref{eq:bloch}, the non-linear term, $\mc{K}$, drastically simplifies. Since $|\braket{0}{\psi_t}|^2=(1+z_t)/2$ and $|\braket{1}{\psi_t}|^2=(1-z_t)/2$, $\mc{K}_t$ can be written as an effectively state-dependent Hamiltonian \cite{Childs2016PRA},
\begin{equation}
\mc{K}=\begin{pmatrix}
\kappa\left(\sqrt{(1+z_t)/2}\right)&0\\
0&\kappa\left(\sqrt{(1-z_t)/2}\right)\,.
\end{pmatrix}
\end{equation}
Observe that $\mc{K}$ is diagonal in the Bloch representation. 

It is then a simple exercise to show that the non-linear Schr\"odinger equation \eqref{eq:nonlin_sch} can be written as
\begin{equation}
\label{eq:diffeq}
\begin{aligned}[b]
\dot{x}_t&=-\left[2\zeta +\tilde{\kappa}(z_t)\right]\,y_t+ 2\xi z_t\\
\dot{y}_t&=\quad\left[2\zeta +\tilde{\kappa}(z_t)\right]x_t-2 \chi z_t\\
\dot{z}_t&=-2\chi x_t + 2 \xi y_t\,.
\end{aligned}
\end{equation}
where $\tilde{\kappa}(z_t)\equiv\kappa(\sqrt{(1+z_t)/2})-\kappa(\sqrt{(1-z_t)/2})$.

\paragraph*{Stationary states}

In the present analysis, we are interested in the thermodynamic properties of nonlinear qubits. The only states that can be fully characterized by means of thermodynamics are stationary, equilibrium states. Therefore, we now continue by determining the stationary solutions of Eq.~\eqref{eq:diffeq},
\begin{equation}
\label{eq:stat}
\begin{aligned}[b]
0&=-\left[2\zeta +\tilde{\kappa}(z)\right]\,y+ 2\xi z\\
0&=\quad\left[2\zeta +\tilde{\kappa}(z)\right]x-2 \chi z\\
0&=-2\chi x + 2 \xi y\,.
\end{aligned}
\end{equation}
We observe that Eqs.~\eqref{eq:stat} are an under-determined system, which can equivalently be written as,
\begin{equation}
x=\frac{2\xi\,z}{2\zeta+\tilde{\kappa}(z)}\quad\text{and}\quad y=\frac{2\chi\,z}{2\zeta+\tilde{\kappa}(z)}\,.
\end{equation}
Thus, for all stationary states the length of the Bloch vector becomes,
\begin{equation}
\label{eq:vec}
|\vec{r}|\equiv r=|z|\,\sqrt{1+\frac{4\, (\xi^2+\chi^2)}{[2 \zeta+\tilde{\kappa}(z)]^2}}\,,
\end{equation}
which still explicitly depends on the $z$ component. In the next section, we will now determine the values of $z$, or rather the stationary quantum states that correspond to thermodynamic equilibrium in the canonical representation.

\section{Canonical equilibrium state}

For any quantum state $\rho$, the energy is given by the expectation value of the Hamiltonian. For the present case, we can write,
\begin{equation}
\label{eq:energy}
E=\tr{H \rho}=\vec{B}\cdot \vec{r} =\frac{2 B^2+ \zeta \tilde{\kappa}(z)}{2 \zeta +\tilde{\kappa}(z)}\,z \,,
\end{equation}
where we introduced $B\equiv |\vec{B}|=\sqrt{\xi^2+\chi^2+\zeta^2}$.

Since we are interested in stationary, equilibrium states, the thermodynamic entropy is given by the corresponding von Neumann entropy, $S=-\tr{\rho \ln{\rho}}$, which only depends on the length of the Bloch vector \eqref{eq:vec},
\begin{equation}
\label{eq:entropy}
S=\ln(2)-\frac{1}{2}\ln\left(1-r^2\right)-r\, \mrm{arctanh}(r)\,,
\end{equation}
where we also set $k_B=1$ to avoid clutter. Note that for arbitrary, nonequilibrrium states, $S$, only quantifies the information content that is related, but identical to the thermodynamic entropy. In the present analysis, we only consider stationary, equilibrium states, and hence the identification \eqref{eq:entropy} is sound.

 As usual, this thermodynamic entropy is maximal in equilibrium. In the canonical representation, we can then determine the quantum state $\rho$ that has the maximal entropy given that the energy \eqref{eq:energy} is known and constant. Employing standard variational calculus we write
\begin{equation}
\label{eq:var}
0=\delta\left(S-\beta E\right)=\delta z\,\left(\frac{\pd S}{\pd z}-\beta \frac{\pd E}{\pd z}\right)\,,
\end{equation}
where $\beta$ is a Langrange multiplier which can be identified with the inverse temperature $\beta=1/T$ \footnote{Note that Eq.~\eqref{eq:var} implies $\beta=\pd S/\pd E$, which is identical to the thermodynamic definition of temperature \cite{Callen1985}.}.

While Eq.~\eqref{eq:var} is formally identical to standard problems in thermodynamics, here the resulting equations are nonlinear. Therefore, analytical solutions can only be found in very special cases and generally one has to rely to numerical treatments.

\paragraph*{Linear qubits}

To demonstrate the principles and build intuition, we briefly review the linear case,i.e., $\tilde{\kappa}(z)=0$. In this case, the internal energy \eqref{eq:energy} becomes
\begin{equation}
\label{eq:energy_lin}
E_\mrm{lin}= \frac{z}{\zeta}\,B^2\,,
\end{equation}
and the length of the Bloch vector \eqref{eq:vec} can be written as
\begin{equation}
r_\mrm{lin}=\frac{|z|}{|\zeta|}\,B\,,
\end{equation}
Correspondingly, the equilibrium state is determined from Eq.~\eqref{eq:var} as a solution of
\begin{equation}
0=\beta B^2 + B\, \mrm{arctanh}\left(B \,z/\zeta\right)\,,
\end{equation}
and hence
\begin{equation}
z=-\frac{\zeta}{B}\,\tanh{\left(\beta B\right)}\,.
\end{equation}
Note that this solution represents nothing but the Gibbs state, $\rho=\exp{(-\beta H})/Z$, and as a result, one finds the known expression of the internal energy,
\begin{equation}
\label{eq:energy_linear}
E_\mrm{lin}= -B\,\tanh{\left(\beta B\right)}
\end{equation}
Note that the sign is determined by the direction of the magnetic field. Shortly, it will also be useful express the stationary Bloch vector as
\begin{equation}
\label{eq:r_lin}
r_\mrm{lin}=|\tanh{\left(\beta B\right)}|\,.
\end{equation}

\paragraph*{Gross-Pitaevskii qubits}

The mathematical treatment becomes significantly more involved for nonlinear qubits. Since we have seen in previous work that different nonlinearities give rise to different details in the dynamics \cite{Deffner2022EPL,Deffner2025QST}, but also that the stationary properties are somewhat generic \cite{Deffner2025EPL}, we will now continue the analysis for a specific choice of $\tilde{\kappa}(z)$. Arguably, the most tractable case is given by nonlinearities of the Gross-Piteaveskii type, $\tilde{\kappa}(z)=g z$. 

In this case, the internal energy \eqref{eq:energy} becomes
\begin{equation}
\label{eq:energy_GPE}
E_\mrm{GPE}=\frac{2 B^2+ \zeta g\,z}{2 \zeta +g\,z}\,z\,,
\end{equation}
and for the Bloch vector we have
\begin{equation}
r_\mrm{GPE}= \frac{|z|}{|2 \zeta+g z|}\,\sqrt{4 B^2+ 4\zeta g z+g^2 z^2}\,.
\end{equation}
Even for this ``most tractable case'', Eq.~\eqref{eq:var} becomes a rather involved, lengthy equation that can no longer be solved in ``neat'' form. It is worth emphasizing again that the resulting stationary state of an arbitrary, nonlinear qubit is \emph{not} the Gibbs state. Nevertheless, the qubit is in thermodynamic equilibrium and its corresponding state is the proper ``thermal state''.

\begin{figure}
\includegraphics[width=.48\textwidth]{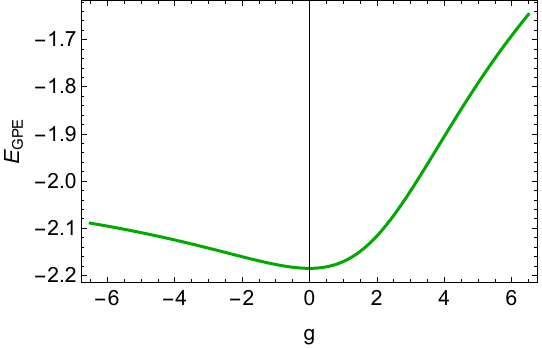}
\caption{\label{fig:energy_g} Internal energy \eqref{eq:energy_GPE} for a Gross-Pitaevskii-type nonlinearity, $\tilde{\kappa}(z)=g z$, as a function of $g$ at $\beta=1$. Parameters are $\xi=1$, $\chi=0$, and $\zeta=2$.}
\end{figure}

In Fig.~\ref{fig:energy_g} we depict the internal energy \eqref{eq:energy_GPE} for a numerical solution of Eq.~\eqref{eq:var} at constant temperature as a function of the strength of the nonlinearity $g$. We observe that $E_\mrm{GPE}$ is significantly increased by the presence of the nonlinearity.

\begin{figure*}
\includegraphics[width=.32\textwidth]{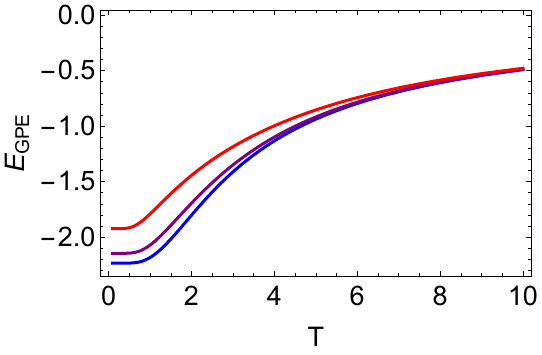}\hfill  \includegraphics[width=.32\textwidth]{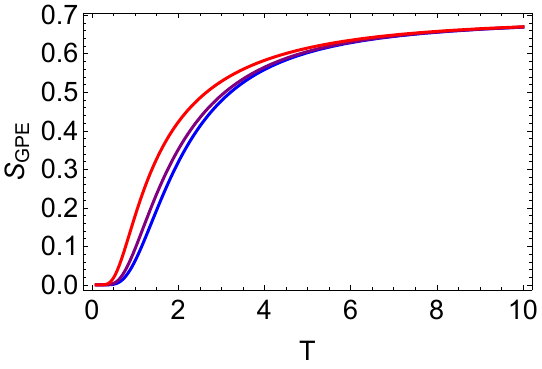}\hfill \includegraphics[width=.32\textwidth]{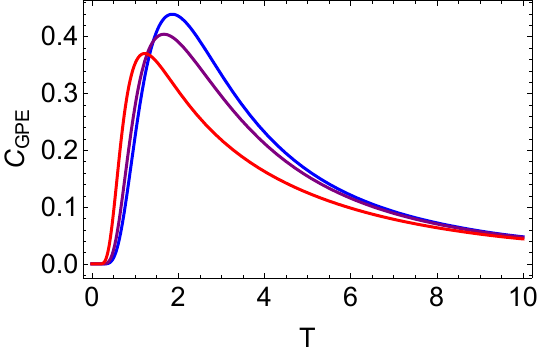}
\caption{\label{fig:energy} Internal energy \eqref{eq:energy_GPE}, entropy \eqref{eq:entropy}, and heat capacity with a Gross-Pitaevskii nonlinearity, $\tilde{\kappa}(z)=g z$, for $g=0$ (blue), $g=2.5$ (purple), and $g=5$ (red) as a function of temperature $T$. Parameters are $\xi=1$, $\chi=0$, and $\zeta=2$.}
\end{figure*}

Furthermore, in Figs.~\ref{fig:energy} we plot the internal energy \eqref{eq:energy_GPE}, the entropy \eqref{eq:entropy}, and the heat capacity $C$ as a function of temperature. Like always $C$ is defined as $C\equiv \pd E/\pd T$. We observe that both energy and entropy are increased in the presence of nonlinearities, and that the effect is the stronger the higher the value of $g$. Interestingly, the maximal value of the corresponding heat capacity is smaller, which indicates that nonlinear qubits can store less heat at the same temperature as linear qubits, while the characteristic ``Schottky hump'' qualitatively survives \cite{Callen1985}.

The natural question now has to be whether this nonlinear enhancement in internal energy can be leveraged in thermal devices. To this end, we continue the discussion by analyzing quantum Otto engines that operate with such nonlinear qubits as a working medium. 

\section{Nonlinear quantum Otto engine}

Among all thermodynamic cycles, the Otto cycle is particularly convenient when analyzing quantum engines \cite{Deffner2019book}. The ideal Otto cycle consists of four strokes, of which two are isentropic and two are isochoric. Thus, during each stroke the engine exchanges either heat or work with the environment, but never both. This also means that the cycle can be fully described in terms of changes in internal energy, and one does not have to worry about the myriad of notions of ``quantum work'' that have been put forward \cite{Deffner2019book,campbell2025roadmap}.

\subsection{Ideal quantum Otto cycle}

We begin with the ideal Otto cycle, in which the qubit, aka the working medium, remains in a state of thermodynamic equilibrium at all times.

\paragraph*{Isentropic compression -- $A\rightarrow B$:} 

During the first stroke of the cycle, the working medium is decoupled from the heat reservoirs and hence no heat can be absorbed, $Q_{A\rightarrow B}=0$ and $S_B=S_A$. Therefore, the work is given by the change of internal energy,
\begin{equation}
\label{eq:work_AB}
W_{A\rightarrow B}= E_B-E_A\,.
\end{equation}
Note that ``compression'' of a qubit corresponds to an increase in the magnetic field, $|B_B|>|B_A|$.

\paragraph*{Isochoric heating -- $B\rightarrow C$:}

During the second, isochoric stroke the magnetic field is held constant, $B_C=B_B$, and hence no work is performed by the engine, $W_{B\rightarrow C}=0$. Instead, the working medium absorbs heat from the hot environment and we can write,
\begin{equation}
Q_{B\rightarrow C}=E_C-E_B\,.
\end{equation}

\paragraph*{Isentropic expansion -- $C\rightarrow D$:}

During the isentropic expansion, the magnetic field returns to its original value, $B_D=B_A$, while the entropy is held constant, $Q_{C\rightarrow D}=0 $ and $S_C=S_D$. Accordingly,
\begin{equation}
W_{C\rightarrow D}= E_C-E_D\,.
\end{equation}

\paragraph*{Isochoric cooling -- $D\rightarrow A$:}

Finally, the cycle is closed by an ischoric cooling stroke with
\begin{equation}
\label{eq:heat_DA}
Q_{D\rightarrow A}=E_D-E_A\,,
\end{equation}
and $W_{D\rightarrow A}=0$.

\paragraph*{Ideal Otto efficiency}

\begin{figure}
\includegraphics[width=.48\textwidth]{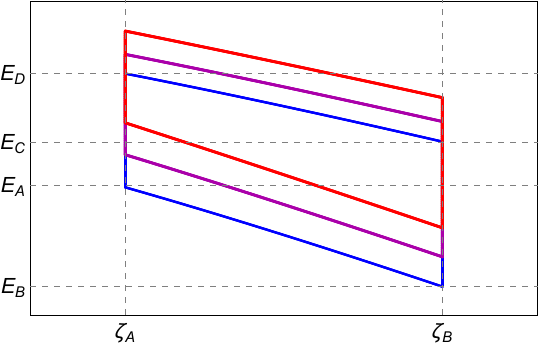}
\caption{\label{fig:Ezeta} $E-\zeta$-diagram for the ideal Otto cycle for a Gross-Pitaevskii nonlinearity, $\tilde\kappa(z)=gz$ for $g=0$ (blue), $g=-2.5$ (purple), and $g=-5$ (red). Parameters are $\xi=1$, $\chi=0$, $\zeta_A=1$, $\zeta_B=1.5$, $\beta_A=1$, and $\beta_D=0.5$.}
\end{figure}

For the sake of simplicity and without loss of generality, we now assume that the magnetic field is varied only in $z$-direction. Then, in Fig.~\ref{fig:Ezeta} we plot the resulting $E-\zeta$-diagram, which plays the role of the more standard pressure-volume diagrams. We observe (as expected) that the diagram is shifted ``upwards'' for nonlinear qubits, which reflects the nonlinear increase in internal energy. 

The more pertinent question is to what extend the efficiency of the cycle is modified in the presence of a nonlinearity. It is easy to see that we have
\begin{equation}
\label{eq:eta}
\eta=-\frac{W_{A\rightarrow B}+W_{C\rightarrow D}}{Q_{B\rightarrow C}}=1-\frac{Q_{D\rightarrow A}}{Q_{B\rightarrow C}}\,,
\end{equation}
which we will now evaluate explicitly for linear and nonlinear engines.

\subsection{Linear Otto engine}

Mostly for pedagogical reasons and for the sake of completeness, we briefly review the quantum Otto efficiency for linear qubits, see also Refs.~\cite{Quan2007PRE}. To this end, note that for isentropic strokes we have $r=\mrm{const.}$, and hence from Eq.~\eqref{eq:r_lin}, $\beta_A B_A=\beta_B B_B$ and $\beta_C B_B=\beta_D B_A$. Thus we can immediately write,
\begin{equation}
\label{eq:eta_lin}
\eta=1-\frac{E_D-E_A}{E_C-E_B}=1-\frac{B_A}{B_B}\equiv 1-\mc{C},
\end{equation}
where we only used Eq.~\eqref{eq:energy_linear}. 

In complete analogy to Otto engines that operate with classical gases \cite{Callen1985,Smith2020JNET}, harmonic oscillators \cite{Abah2012PRL,Deffner2018Entropy,Myers2020PRE,Myers2021PRXQ} or even relativistic oscillators \cite{Myers2021NJP}, the efficiency is given by one minus the compression ratio, $\mc{C}\equiv B_A/B_B$.

\subsection{Ideal Gross-Pitaevskii engine}

\begin{figure}
\includegraphics[width=.48\textwidth]{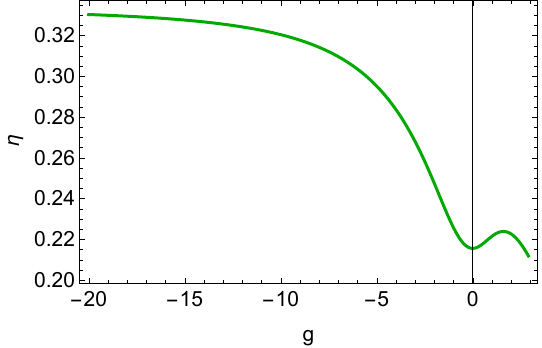}
\caption{\label{fig:eta} Efficiency of the ideal Otto cycle \eqref{eq:eta} for qubits with a Gross-Pitaevskii nonlinearity, $\tilde\kappa(z)=gz$. Parameters are $\xi=1$, $\chi=0$, $\zeta_A=1$, $\zeta_B=1.5$, $\beta_A=1$, and $\beta_D=0.5$.}
\end{figure}

For nonlinear qubits, we again have to employ a numerical solution. To this end, we computed Eq.~\eqref{eq:eta} for a wide range of values for $g$. The result is depicted in Fig.~\ref{fig:eta}.

We immediately observe that the efficiency of nonlinear engines is larger that for the linear case, $g=0$. Moreover, the effect is much more pronounced for repulsive nonlinearities $g<0$. For very large, negative values of $g$, the efficiency saturates. In this regime, the quantum state is dominated by the nonlinearity, and the thermodynamic properties are only weakly dependent on the parameters of the Hamiltonian. Moreover, we observer a similar asymmetry of the efficiency as for the internal energy, cf. \ref{fig:energy_g}. Note, however, that the increase in energy as well as efficiency are given by nonlinear terms, which makes the observed effect very sensitive to the choice of parameters.

\section{Endoreversible Otto cycle -- efficiency at maximum power}

While the above observed increase in ideal efficiency is interesting, we do need to wonder whether it is ``real''. The more practically relevant question is whether also the efficiency at maximal power output is enhanced by the presence of nonlinearities in the working medium. We answer this question within the framework of endoreversible thermodynamics \cite{Hoffmann1997}. In endoreversible thermodynamics one assumes that all processes are slow enough that the system remains in a state of \emph{locally equilibrium}, yet the processes are too fast for the system to reach a state of equilibrium with the environment. More specifically, imagine an engine, whose working medium is at equilibrium at temperature $T$. However, $T$ is not equal to the temperature of the heat bath, $T_\mrm{bath}$, and thus there is a temperature gradient at the boundaries of the engine. Now further imagine that the engine undergoes a slow, cyclic state transformation, where slow means that the working medium remains \emph{locally} in equilibrium at all times. Then, from the point of view of the environment the device undergoes an irreversible cycle. Such state transformations are called \emph{endoreversible} \cite{Hoffmann1997}, which means that locally the transformation is reversible, but globally irreversible. 

In a seminal work, Curzon and Ahlborn showed \cite{Curzon1975AJP} that the efficiency of a Carnot engine undergoing an endoreversible cycle at maximal power is given by,
\begin{equation}
\label{eq:CA}
\eta_\mrm{CA}=1-\sqrt{\frac{T_c}{T_h}}=1-\sqrt{\frac{\beta_h}{\beta_c}}\,,
\end{equation}
where $\beta_c$ and $\beta_h$ are the inverse temperatures of the cold and hot reservoirs, respectively. Since its discovery the Curzon-Ahlborn efficiency \eqref{eq:CA} has received a great deal of attention.  Recently it has been found, e.g., that also endorversible Otto \cite{Deffner2018Entropy} and Brayton \cite{Ferketic2023EPL} engines operating with ideal gases have the same efficiency. However, more generally the efficiency depends on the choice of working medium and the resulting efficiency can be larger \cite{Myers2020PRE,Smith2020JNET,Myers2021Symmetry,Myers2021NJP,Myers2021PRXQ,Myers2022NJP,Myers2023Nanomat,Pena2023Entropy} or smaller \cite{Behrendt2025entropy} than $\eta_\mrm{CA}$.

\subsection{Endoreversible quantum Otto cycle}

In the present analysis, we focus on quantum Otto engines, of which we developed the endoreversible version in Ref.~\cite{Deffner2018Entropy}. This engine cycle has the following four strokes:

\paragraph*{Isentropic compression: $A\rightarrow B$:}

As stated above, during the isentropic strokes the working qubit does not exchange heat with the environment. Therefore, the thermodynamic state of the working qubit can be considered independent of the environment, and the endoreversible description is identical to the equilibrium cycle, see Eq.~\eqref{eq:work_AB}

\paragraph*{Isochoric heating -- $B\rightarrow C$:}

During the isochoric strokes the magnetic field is held constant, and the system exchanges \emph{only} heat with the environment. For endoreversible cycles, we now assume that the working qubit is in a state of local equilibrium, but also that the working qubit never fully equilibrates with the hot reservoir. Therefore, we can write
\begin{equation}
\beta(0)=\beta_B\quad\mrm{and}\quad \beta(\tau_h)=\beta_C\quad\mrm{with}\quad \beta_B>\beta_C \geq \beta_h\,,
\end{equation}
where as before $\tau_h$ is the duration of the stroke.

The change in temperature from $\beta_B$ to $\beta_C$ is governed by Fourier's law \cite{Callen1985},
\begin{equation}
\label{eq:fourier_hot}
\frac{d \beta}{dt}=\alpha_h \left(\beta(t)-\beta_h\right)
\end{equation}
where $\alpha_h$ is a constant depending on the heat conductivity and heat capacity of the working qubit. 

Equation~\eqref{eq:fourier_hot} can be solved analytically, and we have
\begin{equation}
\label{eq:rel_hot}
\beta_C-\beta_h=\left(\beta_B-\beta_h\right)\,\e{\alpha_h \tau_h}\,.
\end{equation}

\paragraph*{Isentropic expansion -- $C\rightarrow D$:} 

Again, this stroke is identical to the ideal case, since the working medium is not in contact with the thermal environment.

\paragraph*{Isochoric cooling -- $D\rightarrow A$:}

During isochoric cooling, the work again vanishes, and the heat is given by the change in internal energy \eqref{eq:heat_DA}. However, once again we have to account for the thermal gradient between working medium and heat reservoir. Thus, we write now,
\begin{equation}
\beta(0)=\beta_D\quad\mrm{and}\quad \beta(\tau_c)=\beta_A\quad\mrm{with}\quad \beta_D<\beta_A \leq \beta_c\,.
\end{equation}

In complete analogy to above \eqref{eq:fourier_hot}, the heat transfer is described by Fourier's law,
\begin{equation}
\label{eq:fourier_cold} 
\frac{d \beta}{dt}=\alpha_c \left(\beta(t)-\beta_c\right)\,,
\end{equation}
where $\alpha_c$ is a constant characteristic for the cold stroke. As solution of Eq.~\eqref{eq:fourier_cold} we now obtain,
\begin{equation}
\label{eq:rel_cold}
\beta_A-\beta_c=\left(\beta_D-\beta_c\right)\,\e{\alpha_c \tau_c}\,,
\end{equation}
which properly describes the decrease in temperature from $\beta_D$ back to $\beta_A$.

\subsection{Linear qubit engine}

\begin{figure}
\includegraphics[width=.48\textwidth]{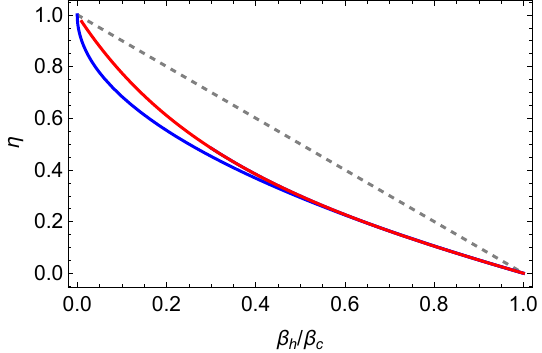}
\caption{\label{fig:eta_linear} Efficiency at maximum power of the endoreversible Otto cycle for linear qubits (red), together with the Curzon-Ahlborn efficiency \eqref{eq:CA} (blue) and the Carnot efficiency (gray dashed). Parameters are $\xi=1$, $\chi=0$, $\alpha=1$, $\tau=1$, and $\zeta_A=1$. }
\end{figure}

We are now interested in the efficiency $\eta$ at which the power output, 
\begin{equation}
P\equiv -\frac{W_{A\rightarrow B}+W_{C\rightarrow D}}{\gamma (\tau_h+\tau_c)}\,,
\end{equation}
is maximal. As always, the cycle time is given as a multiple of the strokes times during which the working medium is in contact with the heat reservoirs \cite{Deffner2018Entropy}, $\tau_\mrm{cyc}=\gamma(\tau_h+\tau_c)$.

To simplify the expressions we set $\alpha_h=\alpha_c=\alpha$ and $\tau_h=\tau_c=\tau$. Further using again $\beta_A B_A=\beta_B B_B$ and $\beta_C B_B=\beta_D B_A$, we can write
\begin{equation}
\label{eq:power}
P=\frac{\mc{C}-1}{\mc{C}} \frac{B_A}{2 \gamma \tau} \left[\mrm{tanh}\left(\beta_A B_A\right)-\mrm{tanh}\left(\beta_D B_A\right)\right]\,,
\end{equation}
which still depends on the qubit temperatures $\beta_A$ and $\beta_D$. These temperatures can be eliminated employing Eqs.~\eqref{eq:rel_hot} and \eqref{eq:rel_cold}, from which we obtain,
\begin{equation}
\beta_A=\frac{e^{\alpha \tau}\,\beta_h+\mc{C}\,\beta_c }{\mc{C}\,(1+e^{\alpha \tau})}\quad\mrm{and}\quad\beta_D=\frac{\beta_h+\mc{C}\,e^{\alpha \tau}\,\beta_c }{\mc{C}\,(1+e^{\alpha \tau})}\,.
\end{equation}
Substituting the latter into Eq.~\eqref{eq:power} leads to a somewhat lengthy expression for the power $P$ as a function of the compression ratio $\mc{C}$, and the hot and cold temperatures, $\beta_h$ and $\beta_c$.

The resulting expression can then be maximized as a function of $\mc{C}$, from which we obtain the efficiency at maximum power with Eq.~\eqref{eq:eta_lin}. The result is plotted in Fig.~\ref{fig:eta_linear}. We observe that quantum Otto engines operating in the endoreversible regime with linear qubits as working medium are more efficient than the Curzon-Ahlborn efficiency \eqref{eq:CA}. This result corroborates our earlier findings showing that only working mediums with caloric equations of state that are linear in temperature lead to the Curzon-Ahlborn efficiency \cite{Smith2020JNET}.

\subsection{Endoreversible Gross-Pitaevskii engine}

To conclude the analysis, we now also determine the efficiency at maximum power of an Otto engine with Gross-Piteavskii qubits as working medium. Solving this problem in full generality is a mathematically very involved problem. First, the adiabaticity relations have to be solved, $S_A=S_B$ and $S_C=S_D$, from which we can express $\beta_A(\beta_B;\,B_A,B_B)$ and $\beta_C(\beta_D;\,B_A,B_B)$. From that, we then need to determine the $z$-component of the equilibrium state by solving Eq.~\eqref{eq:var}, from which we then can obtain the nonlinear generalization of expression \eqref{eq:power}. Since all of this has to be done numerically, the gained insight is somewhat limited. 

More direct insight can be gained for weak nonlinarities, $g\ll 1$, for which we can expand the Eq.~\eqref{eq:energy_GPE} for the internal energy. We have,
\begin{equation}
\label{eq:energy_approx}
E_\mrm{GPE}\simeq  \frac{z}{\zeta}\,B^2-g\,\frac{z^2\left(\xi^2+\chi^2\right)}{\zeta^2}+\mc{O}(g^2)\,.
\end{equation}
Correspondingly, we can determine the correction to the power \eqref{eq:power} in linear order of $g$, arising from the Gross-Pitaevskii nonlinearity. The resulting expression does become rather lengthy, but tractable.

\begin{figure}
\includegraphics[width=.48\textwidth]{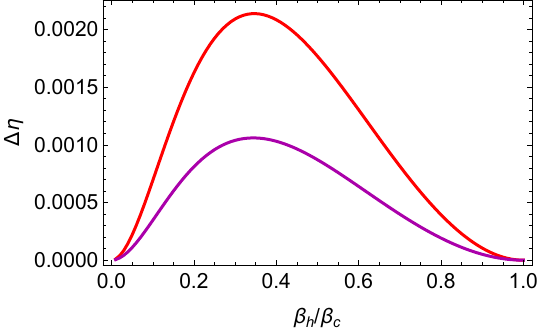}
\caption{\label{fig:eta_delta} Difference in efficiency at maximum power of the endoreversible Otto cycle \eqref{eq:etadiff} with a Gross-Pitaevskii nonlinearity, $\tilde\kappa(z)=gz$, for $g=0.05$ (purple) and $g=0.1$ (red). Parameters are $\xi=1$, $\chi=0$, $\alpha=1$, $\tau=1$, and $\zeta_A=1$.}
\end{figure}

In Fig.~\ref{fig:eta_delta} we plot the difference of the efficiency at maximum power for Otto engines running with Gross-Piteaveskii and linear qubits,
\begin{equation}
\label{eq:etadiff}
\Delta\eta=\eta_\mrm{GPE}-\eta_\mrm{linear}\,.
\end{equation}
For the numerical solution, we worked in the regime of small $g$ with Eq.~\eqref{eq:energy_approx}. We observe that in complete analogy to the ideal case, cf. Fig.~\ref{fig:eta}, the efficiency for nonlinear qubits is higher. However, we also observe that the nonlinear enhancement is a nontrivial function, which as a maximum at intermediate temperature ratios $\beta_h/\beta_c$.

\section{Concluding remarks}

In the present analysis we have found that thermodynamic devices built with nonlinear qubits can significantly outperform linear devices. To this end, we have developed a comprehensive thermodynamics of general nonlinear qubits, including the quantum state corresponding to canonical equilibrium, internal energy, entropy, heat capacity, as well as the efficiency of ideal and endoreversible Otto cycles. In particular, for the quantum Otto engine we found that nonlinear device can operate at significantly enhanced efficiency.

The present findings are consistent and complementary to our previous work. We have now found that nonlinear systems evolve faster \cite{Deffner2022EPL}, permit higher precision in measurements \cite{Deffner2025QST}, suppress parasitic excitations \cite{Deffner2025EPL}, and have better thermodynamic performance.

\acknowledgments{S.D. acknowledges support from the John Templeton Foundation under Grant No. 63626. This work was supported by the U.S. Department of Energy, Office of Basic Energy Sciences, Quantum Information Science program in Chemical Sciences, Geosciences, and Biosciences, under Award No. DE-SC0025997.}

\bibliography{nonlinear_thermo}

\begin{thebibliography}{64}%
\makeatletter
\providecommand \@ifxundefined [1]{%
 \@ifx{#1\undefined}
}%
\providecommand \@ifnum [1]{%
 \ifnum #1\expandafter \@firstoftwo
 \else \expandafter \@secondoftwo
 \fi
}%
\providecommand \@ifx [1]{%
 \ifx #1\expandafter \@firstoftwo
 \else \expandafter \@secondoftwo
 \fi
}%
\providecommand \natexlab [1]{#1}%
\providecommand \enquote  [1]{``#1''}%
\providecommand \bibnamefont  [1]{#1}%
\providecommand \bibfnamefont [1]{#1}%
\providecommand \citenamefont [1]{#1}%
\providecommand \href@noop [0]{\@secondoftwo}%
\providecommand \href [0]{\begingroup \@sanitize@url \@href}%
\providecommand \@href[1]{\@@startlink{#1}\@@href}%
\providecommand \@@href[1]{\endgroup#1\@@endlink}%
\providecommand \@sanitize@url [0]{\catcode `\\12\catcode `\$12\catcode
  `\&12\catcode `\#12\catcode `\^12\catcode `\_12\catcode `\%12\relax}%
\providecommand \@@startlink[1]{}%
\providecommand \@@endlink[0]{}%
\providecommand \url  [0]{\begingroup\@sanitize@url \@url }%
\providecommand \@url [1]{\endgroup\@href {#1}{\urlprefix }}%
\providecommand \urlprefix  [0]{URL }%
\providecommand \Eprint [0]{\href }%
\providecommand \doibase [0]{https://doi.org/}%
\providecommand \selectlanguage [0]{\@gobble}%
\providecommand \bibinfo  [0]{\@secondoftwo}%
\providecommand \bibfield  [0]{\@secondoftwo}%
\providecommand \translation [1]{[#1]}%
\providecommand \BibitemOpen [0]{}%
\providecommand \bibitemStop [0]{}%
\providecommand \bibitemNoStop [0]{.\EOS\space}%
\providecommand \EOS [0]{\spacefactor3000\relax}%
\providecommand \BibitemShut  [1]{\csname bibitem#1\endcsname}%
\let\auto@bib@innerbib\@empty
\bibitem [{\citenamefont {Campbell}\ \emph {et~al.}(1985)\citenamefont
  {Campbell}, \citenamefont {Farmer}, \citenamefont {Crutchfield},\ and\
  \citenamefont {Jen}}]{Campbell1985CACM}%
  \BibitemOpen
  \bibfield  {author} {\bibinfo {author} {\bibfnamefont {D.}~\bibnamefont
  {Campbell}}, \bibinfo {author} {\bibfnamefont {D.}~\bibnamefont {Farmer}},
  \bibinfo {author} {\bibfnamefont {J.}~\bibnamefont {Crutchfield}},\ and\
  \bibinfo {author} {\bibfnamefont {E.}~\bibnamefont {Jen}},\ }\bibfield
  {title} {\bibinfo {title} {Experimental mathematics: the role of computation
  in nonlinear science},\ }\href {https://doi.org/10.1145/3341.3345} {\bibfield
   {journal} {\bibinfo  {journal} {Commun. ACM}\ }\textbf {\bibinfo {volume}
  {28}},\ \bibinfo {pages} {374} (\bibinfo {year} {1985})}\BibitemShut
  {NoStop}%
\bibitem [{\citenamefont {Jordan}(2009)}]{Jordan2009JPCF}%
  \BibitemOpen
  \bibfield  {author} {\bibinfo {author} {\bibfnamefont {T.~F.}\ \bibnamefont
  {Jordan}},\ }\bibfield  {title} {\bibinfo {title} {Why quantum dynamics is
  linear},\ }\href {https://doi.org/10.1088/1742-6596/196/1/012010} {\bibfield
  {journal} {\bibinfo  {journal} {J. Phys.: Conf. Ser.}\ }\textbf {\bibinfo
  {volume} {196}},\ \bibinfo {pages} {012010} (\bibinfo {year}
  {2009})}\BibitemShut {NoStop}%
\bibitem [{\citenamefont {Walsworth}\ \emph {et~al.}(1990)\citenamefont
  {Walsworth}, \citenamefont {Silvera}, \citenamefont {Mattison},\ and\
  \citenamefont {Vessot}}]{Walsworth1990PRL}%
  \BibitemOpen
  \bibfield  {author} {\bibinfo {author} {\bibfnamefont {R.~L.}\ \bibnamefont
  {Walsworth}}, \bibinfo {author} {\bibfnamefont {I.~F.}\ \bibnamefont
  {Silvera}}, \bibinfo {author} {\bibfnamefont {E.~M.}\ \bibnamefont
  {Mattison}},\ and\ \bibinfo {author} {\bibfnamefont {R.~F.~C.}\ \bibnamefont
  {Vessot}},\ }\bibfield  {title} {\bibinfo {title} {Test of the linearity of
  quantum mechanics in an atomic system with a hydrogen maser},\ }\href
  {https://doi.org/10.1103/PhysRevLett.64.2599} {\bibfield  {journal} {\bibinfo
   {journal} {Phys. Rev. Lett.}\ }\textbf {\bibinfo {volume} {64}},\ \bibinfo
  {pages} {2599} (\bibinfo {year} {1990})}\BibitemShut {NoStop}%
\bibitem [{\citenamefont {Majumder}\ \emph {et~al.}(1990)\citenamefont
  {Majumder}, \citenamefont {Venema}, \citenamefont {Lamoreaux}, \citenamefont
  {Heckel},\ and\ \citenamefont {Fortson}}]{Majumder1990PRL}%
  \BibitemOpen
  \bibfield  {author} {\bibinfo {author} {\bibfnamefont {P.~K.}\ \bibnamefont
  {Majumder}}, \bibinfo {author} {\bibfnamefont {B.~J.}\ \bibnamefont
  {Venema}}, \bibinfo {author} {\bibfnamefont {S.~K.}\ \bibnamefont
  {Lamoreaux}}, \bibinfo {author} {\bibfnamefont {B.~R.}\ \bibnamefont
  {Heckel}},\ and\ \bibinfo {author} {\bibfnamefont {E.~N.}\ \bibnamefont
  {Fortson}},\ }\bibfield  {title} {\bibinfo {title} {Test of the linearity of
  quantum mechanics in optically pumped $^{201}\mathrm{Hg}$},\ }\href
  {https://doi.org/10.1103/PhysRevLett.65.2931} {\bibfield  {journal} {\bibinfo
   {journal} {Phys. Rev. Lett.}\ }\textbf {\bibinfo {volume} {65}},\ \bibinfo
  {pages} {2931} (\bibinfo {year} {1990})}\BibitemShut {NoStop}%
\bibitem [{\citenamefont {Forstner}\ \emph {et~al.}(2020)\citenamefont
  {Forstner}, \citenamefont {Zych}, \citenamefont {Basiri-Esfahani},
  \citenamefont {Khosla},\ and\ \citenamefont {Bowen}}]{Forstner2020Optica}%
  \BibitemOpen
  \bibfield  {author} {\bibinfo {author} {\bibfnamefont {S.}~\bibnamefont
  {Forstner}}, \bibinfo {author} {\bibfnamefont {M.}~\bibnamefont {Zych}},
  \bibinfo {author} {\bibfnamefont {S.}~\bibnamefont {Basiri-Esfahani}},
  \bibinfo {author} {\bibfnamefont {K.~E.}\ \bibnamefont {Khosla}},\ and\
  \bibinfo {author} {\bibfnamefont {W.~P.}\ \bibnamefont {Bowen}},\ }\bibfield
  {title} {\bibinfo {title} {Nanomechanical test of quantum linearity},\ }\href
  {https://doi.org/10.1364/OPTICA.391671} {\bibfield  {journal} {\bibinfo
  {journal} {Optica}\ }\textbf {\bibinfo {volume} {7}},\ \bibinfo {pages}
  {1427} (\bibinfo {year} {2020})}\BibitemShut {NoStop}%
\bibitem [{\citenamefont {DalFavero}\ \emph {et~al.}(2024)\citenamefont
  {DalFavero}, \citenamefont {Meill}, \citenamefont {Meyer}, \citenamefont
  {Wong},\ and\ \citenamefont {Wrubel}}]{DalFavero2024PRA}%
  \BibitemOpen
  \bibfield  {author} {\bibinfo {author} {\bibfnamefont {B.}~\bibnamefont
  {DalFavero}}, \bibinfo {author} {\bibfnamefont {A.}~\bibnamefont {Meill}},
  \bibinfo {author} {\bibfnamefont {D.~A.}\ \bibnamefont {Meyer}}, \bibinfo
  {author} {\bibfnamefont {T.~G.}\ \bibnamefont {Wong}},\ and\ \bibinfo
  {author} {\bibfnamefont {J.~P.}\ \bibnamefont {Wrubel}},\ }\bibfield  {title}
  {\bibinfo {title} {Constant-time quantum search with a many-body quantum
  system},\ }\href {https://doi.org/10.1103/PhysRevA.110.052411} {\bibfield
  {journal} {\bibinfo  {journal} {Phys. Rev. A}\ }\textbf {\bibinfo {volume}
  {110}},\ \bibinfo {pages} {052411} (\bibinfo {year} {2024})}\BibitemShut
  {NoStop}%
\bibitem [{\citenamefont {Gross}(1961)}]{Gross1961}%
  \BibitemOpen
  \bibfield  {author} {\bibinfo {author} {\bibfnamefont {E.~P.}\ \bibnamefont
  {Gross}},\ }\bibfield  {title} {\bibinfo {title} {{Structure of a quantized
  vortex in boson systems}},\ }\href
  {https://link.springer.com/article/10.1007/BF02731494} {\bibfield  {journal}
  {\bibinfo  {journal} {Nuovo Cim.}\ }\textbf {\bibinfo {volume} {20}},\
  \bibinfo {pages} {454} (\bibinfo {year} {1961})}\BibitemShut {NoStop}%
\bibitem [{\citenamefont {Pitaevskii}(1961)}]{Pitaevskii1961}%
  \BibitemOpen
  \bibfield  {author} {\bibinfo {author} {\bibfnamefont {L.~P.}\ \bibnamefont
  {Pitaevskii}},\ }\bibfield  {title} {\bibinfo {title} {{Vortex lines in an
  imperfect Bose gas}},\ }\href
  {http://www.jetp.ras.ru/cgi-bin/e/index/e/13/2/p451?a=list} {\bibfield
  {journal} {\bibinfo  {journal} {Sov. J. Exp. Theor. Phys.}\ }\textbf
  {\bibinfo {volume} {13}},\ \bibinfo {pages} {451} (\bibinfo {year}
  {1961})}\BibitemShut {NoStop}%
\bibitem [{\citenamefont {Rand}(2010)}]{Rand2010}%
  \BibitemOpen
  \bibfield  {author} {\bibinfo {author} {\bibfnamefont {S.}~\bibnamefont
  {Rand}},\ }\href
  {http://books.google.com/books?hl=en{\&}lr={\&}id=dMKc6N9gVs4C{\&}oi=fnd{\&}pg=PR9{\&}dq=Nonlinear+and+Quantum+Optics{\&}ots=uHnGn43Dr{\_}{\&}sig=1tc7diqscyqaccefDaCIV4L2Njk}
  {\emph {\bibinfo {title} {{Nonlinear and Quantum Optics using the density
  matrix}}}}\ (\bibinfo  {publisher} {Oxford University Press},\ \bibinfo
  {year} {2010})\BibitemShut {NoStop}%
\bibitem [{\citenamefont {Ruderman}(2002)}]{Ruderman2002}%
  \BibitemOpen
  \bibfield  {author} {\bibinfo {author} {\bibfnamefont {M.~S.}\ \bibnamefont
  {Ruderman}},\ }\bibfield  {title} {\bibinfo {title} {{Propagation of solitons
  of the Derivative Nonlinear Schr\"odinger equation in a plasma with
  fluctuating density}},\ }\href {https://doi.org/10.1063/1.1482764} {\bibfield
   {journal} {\bibinfo  {journal} {Phys. Plasmas}\ }\textbf {\bibinfo {volume}
  {9}},\ \bibinfo {pages} {2940} (\bibinfo {year} {2002})}\BibitemShut
  {NoStop}%
\bibitem [{\citenamefont {Soler}(1970)}]{Soler1970PRD}%
  \BibitemOpen
  \bibfield  {author} {\bibinfo {author} {\bibfnamefont {M.}~\bibnamefont
  {Soler}},\ }\bibfield  {title} {\bibinfo {title} {Classical, stable,
  nonlinear spinor field with positive rest energy},\ }\href
  {https://doi.org/10.1103/PhysRevD.1.2766} {\bibfield  {journal} {\bibinfo
  {journal} {Phys. Rev. D}\ }\textbf {\bibinfo {volume} {1}},\ \bibinfo {pages}
  {2766} (\bibinfo {year} {1970})}\BibitemShut {NoStop}%
\bibitem [{\citenamefont {Cooper}\ \emph {et~al.}(2010)\citenamefont {Cooper},
  \citenamefont {Khare}, \citenamefont {Mihaila},\ and\ \citenamefont
  {Saxena}}]{Cooper2010PRE}%
  \BibitemOpen
  \bibfield  {author} {\bibinfo {author} {\bibfnamefont {F.}~\bibnamefont
  {Cooper}}, \bibinfo {author} {\bibfnamefont {A.}~\bibnamefont {Khare}},
  \bibinfo {author} {\bibfnamefont {B.}~\bibnamefont {Mihaila}},\ and\ \bibinfo
  {author} {\bibfnamefont {A.}~\bibnamefont {Saxena}},\ }\bibfield  {title}
  {\bibinfo {title} {Solitary waves in the nonlinear dirac equation with
  arbitrary nonlinearity},\ }\href {https://doi.org/10.1103/PhysRevE.82.036604}
  {\bibfield  {journal} {\bibinfo  {journal} {Phys. Rev. E}\ }\textbf {\bibinfo
  {volume} {82}},\ \bibinfo {pages} {036604} (\bibinfo {year}
  {2010})}\BibitemShut {NoStop}%
\bibitem [{\citenamefont {Mertens}\ \emph {et~al.}(2012)\citenamefont
  {Mertens}, \citenamefont {Quintero}, \citenamefont {Cooper}, \citenamefont
  {Khare},\ and\ \citenamefont {Saxena}}]{Mertens2012PRE}%
  \BibitemOpen
  \bibfield  {author} {\bibinfo {author} {\bibfnamefont {F.~G.}\ \bibnamefont
  {Mertens}}, \bibinfo {author} {\bibfnamefont {N.~R.}\ \bibnamefont
  {Quintero}}, \bibinfo {author} {\bibfnamefont {F.}~\bibnamefont {Cooper}},
  \bibinfo {author} {\bibfnamefont {A.}~\bibnamefont {Khare}},\ and\ \bibinfo
  {author} {\bibfnamefont {A.}~\bibnamefont {Saxena}},\ }\bibfield  {title}
  {\bibinfo {title} {Nonlinear dirac equation solitary waves in external
  fields},\ }\href {https://doi.org/10.1103/PhysRevE.86.046602} {\bibfield
  {journal} {\bibinfo  {journal} {Phys. Rev. E}\ }\textbf {\bibinfo {volume}
  {86}},\ \bibinfo {pages} {046602} (\bibinfo {year} {2012})}\BibitemShut
  {NoStop}%
\bibitem [{\citenamefont {Kolomeisky}\ \emph {et~al.}(2000)\citenamefont
  {Kolomeisky}, \citenamefont {Newman}, \citenamefont {Straley},\ and\
  \citenamefont {Qi}}]{Kolomeisky2000PRL}%
  \BibitemOpen
  \bibfield  {author} {\bibinfo {author} {\bibfnamefont {E.~B.}\ \bibnamefont
  {Kolomeisky}}, \bibinfo {author} {\bibfnamefont {T.~J.}\ \bibnamefont
  {Newman}}, \bibinfo {author} {\bibfnamefont {J.~P.}\ \bibnamefont
  {Straley}},\ and\ \bibinfo {author} {\bibfnamefont {X.}~\bibnamefont {Qi}},\
  }\bibfield  {title} {\bibinfo {title} {Low-dimensional bose liquids: Beyond
  the gross-pitaevskii approximation},\ }\href {10.1103/PhysRevLett.85.1146}
  {\bibfield  {journal} {\bibinfo  {journal} {Phys. Rev. Lett.}\ }\textbf
  {\bibinfo {volume} {85}},\ \bibinfo {pages} {1146} (\bibinfo {year}
  {2000})}\BibitemShut {NoStop}%
\bibitem [{\citenamefont {Meyer}\ and\ \citenamefont
  {Wong}(2014)}]{Meyer2014PRA}%
  \BibitemOpen
  \bibfield  {author} {\bibinfo {author} {\bibfnamefont {D.~A.}\ \bibnamefont
  {Meyer}}\ and\ \bibinfo {author} {\bibfnamefont {T.~G.}\ \bibnamefont
  {Wong}},\ }\bibfield  {title} {\bibinfo {title} {Quantum search with general
  nonlinearities},\ }\href {https://doi.org/10.1103/PhysRevA.89.012312}
  {\bibfield  {journal} {\bibinfo  {journal} {Phys. Rev. A}\ }\textbf {\bibinfo
  {volume} {89}},\ \bibinfo {pages} {012312} (\bibinfo {year}
  {2014})}\BibitemShut {NoStop}%
\bibitem [{\citenamefont {Childs}\ and\ \citenamefont
  {Young}(2016)}]{Childs2016PRA}%
  \BibitemOpen
  \bibfield  {author} {\bibinfo {author} {\bibfnamefont {A.~M.}\ \bibnamefont
  {Childs}}\ and\ \bibinfo {author} {\bibfnamefont {J.}~\bibnamefont {Young}},\
  }\bibfield  {title} {\bibinfo {title} {Optimal state discrimination and
  unstructured search in nonlinear quantum mechanics},\ }\href
  {https://doi.org/10.1103/PhysRevA.93.022314} {\bibfield  {journal} {\bibinfo
  {journal} {Phys. Rev. A}\ }\textbf {\bibinfo {volume} {93}},\ \bibinfo
  {pages} {022314} (\bibinfo {year} {2016})}\BibitemShut {NoStop}%
\bibitem [{\citenamefont {Zhang}\ \emph {et~al.}(2020)\citenamefont {Zhang},
  \citenamefont {Wang}, \citenamefont {Ma},\ and\ \citenamefont
  {Ren}}]{Zhang2020MPLB}%
  \BibitemOpen
  \bibfield  {author} {\bibinfo {author} {\bibfnamefont {X.}~\bibnamefont
  {Zhang}}, \bibinfo {author} {\bibfnamefont {C.}~\bibnamefont {Wang}},
  \bibinfo {author} {\bibfnamefont {J.}~\bibnamefont {Ma}},\ and\ \bibinfo
  {author} {\bibfnamefont {G.}~\bibnamefont {Ren}},\ }\bibfield  {title}
  {\bibinfo {title} {Control and synchronization in nonlinear circuits by using
  a thermistor},\ }\href {https://doi.org/10.1142/S021798492050267X} {\bibfield
   {journal} {\bibinfo  {journal} {Mod. Phys. Lett. B}\ }\textbf {\bibinfo
  {volume} {34}},\ \bibinfo {pages} {2050267} (\bibinfo {year}
  {2020})}\BibitemShut {NoStop}%
\bibitem [{\citenamefont {Shen}\ \emph {et~al.}(2023)\citenamefont {Shen},
  \citenamefont {Mok}, \citenamefont {Noh}, \citenamefont {Liu}, \citenamefont
  {Kwek}, \citenamefont {Fan},\ and\ \citenamefont {Chia}}]{Shen2023PRA}%
  \BibitemOpen
  \bibfield  {author} {\bibinfo {author} {\bibfnamefont {Y.}~\bibnamefont
  {Shen}}, \bibinfo {author} {\bibfnamefont {W.-K.}\ \bibnamefont {Mok}},
  \bibinfo {author} {\bibfnamefont {C.}~\bibnamefont {Noh}}, \bibinfo {author}
  {\bibfnamefont {A.~Q.}\ \bibnamefont {Liu}}, \bibinfo {author} {\bibfnamefont
  {L.-C.}\ \bibnamefont {Kwek}}, \bibinfo {author} {\bibfnamefont
  {W.}~\bibnamefont {Fan}},\ and\ \bibinfo {author} {\bibfnamefont
  {A.}~\bibnamefont {Chia}},\ }\bibfield  {title} {\bibinfo {title} {Quantum
  synchronization effects induced by strong nonlinearities},\ }\href
  {https://doi.org/10.1103/PhysRevA.107.053713} {\bibfield  {journal} {\bibinfo
   {journal} {Phys. Rev. A}\ }\textbf {\bibinfo {volume} {107}},\ \bibinfo
  {pages} {053713} (\bibinfo {year} {2023})}\BibitemShut {NoStop}%
\bibitem [{\citenamefont {de~Lacy}\ \emph {et~al.}(2018)\citenamefont
  {de~Lacy}, \citenamefont {Noakes}, \citenamefont {Twamley},\ and\
  \citenamefont {Wang}}]{Lacy2018QIP}%
  \BibitemOpen
  \bibfield  {author} {\bibinfo {author} {\bibfnamefont {K.}~\bibnamefont
  {de~Lacy}}, \bibinfo {author} {\bibfnamefont {L.}~\bibnamefont {Noakes}},
  \bibinfo {author} {\bibfnamefont {J.}~\bibnamefont {Twamley}},\ and\ \bibinfo
  {author} {\bibfnamefont {J.~B.}\ \bibnamefont {Wang}},\ }\bibfield  {title}
  {\bibinfo {title} {Controlled quantum search},\ }\href
  {https://doi.org/10.1007/s11128-018-2031-6} {\bibfield  {journal} {\bibinfo
  {journal} {Quantum Inf. Process.}\ }\textbf {\bibinfo {volume} {17}},\
  \bibinfo {pages} {266} (\bibinfo {year} {2018})}\BibitemShut {NoStop}%
\bibitem [{\citenamefont {Chiew}\ \emph {et~al.}(2019)\citenamefont {Chiew},
  \citenamefont {de~Lacy}, \citenamefont {Yu}, \citenamefont {Marsh},\ and\
  \citenamefont {Wang}}]{Chiew2019QIP}%
  \BibitemOpen
  \bibfield  {author} {\bibinfo {author} {\bibfnamefont {M.}~\bibnamefont
  {Chiew}}, \bibinfo {author} {\bibfnamefont {K.}~\bibnamefont {de~Lacy}},
  \bibinfo {author} {\bibfnamefont {C.~H.}\ \bibnamefont {Yu}}, \bibinfo
  {author} {\bibfnamefont {S.}~\bibnamefont {Marsh}},\ and\ \bibinfo {author}
  {\bibfnamefont {J.~B.}\ \bibnamefont {Wang}},\ }\bibfield  {title} {\bibinfo
  {title} {Graph comparison via nonlinear quantum search},\ }\href
  {https://doi.org/10.1007/s11128-019-2407-2} {\bibfield  {journal} {\bibinfo
  {journal} {Quantum Inf. Process.}\ }\textbf {\bibinfo {volume} {18}},\
  \bibinfo {pages} {302} (\bibinfo {year} {2019})}\BibitemShut {NoStop}%
\bibitem [{\citenamefont {Holmes}\ \emph {et~al.}(2023)\citenamefont {Holmes},
  \citenamefont {Coble}, \citenamefont {Sornborger},\ and\ \citenamefont
  {Subasi}}]{Holmes2023PRA}%
  \BibitemOpen
  \bibfield  {author} {\bibinfo {author} {\bibfnamefont {Z.}~\bibnamefont
  {Holmes}}, \bibinfo {author} {\bibfnamefont {N.~J.}\ \bibnamefont {Coble}},
  \bibinfo {author} {\bibfnamefont {A.~T.}\ \bibnamefont {Sornborger}},\ and\
  \bibinfo {author} {\bibfnamefont {Y.}~\bibnamefont {Subasi}},\ }\bibfield
  {title} {\bibinfo {title} {Nonlinear transformations in quantum
  computation},\ }\href {https://doi.org/10.1103/PhysRevResearch.5.013105}
  {\bibfield  {journal} {\bibinfo  {journal} {Phys. Rev. Res.}\ }\textbf
  {\bibinfo {volume} {5}},\ \bibinfo {pages} {013105} (\bibinfo {year}
  {2023})}\BibitemShut {NoStop}%
\bibitem [{\citenamefont {Geller}(2023{\natexlab{a}})}]{Geller2023CTP}%
  \BibitemOpen
  \bibfield  {author} {\bibinfo {author} {\bibfnamefont {M.~R.}\ \bibnamefont
  {Geller}},\ }\bibfield  {title} {\bibinfo {title} {Nonlinear and non-cp gates
  for bloch vector amplification},\ }\href
  {https://doi.org/10.1088/1572-9494/acf304} {\bibfield  {journal} {\bibinfo
  {journal} {Commun. Theor. Phys.}\ }\textbf {\bibinfo {volume} {75}},\
  \bibinfo {pages} {105102} (\bibinfo {year} {2023}{\natexlab{a}})}\BibitemShut
  {NoStop}%
\bibitem [{\citenamefont {Geller}(2023{\natexlab{b}})}]{Geller2023AQT}%
  \BibitemOpen
  \bibfield  {author} {\bibinfo {author} {\bibfnamefont {M.~R.}\ \bibnamefont
  {Geller}},\ }\bibfield  {title} {\bibinfo {title} {Fast quantum state
  discrimination with nonlinear positive trace-preserving channels},\ }\href
  {https://doi.org/10.1002/qute.202200156} {\bibfield  {journal} {\bibinfo
  {journal} {Adv. Quantum Technol.}\ }\textbf {\bibinfo {volume} {6}},\
  \bibinfo {pages} {2200156} (\bibinfo {year}
  {2023}{\natexlab{b}})}\BibitemShut {NoStop}%
\bibitem [{\citenamefont {Deiml}\ and\ \citenamefont
  {Peterseim}(2024)}]{Deiml2024arXiv}%
  \BibitemOpen
  \bibfield  {author} {\bibinfo {author} {\bibfnamefont {M.}~\bibnamefont
  {Deiml}}\ and\ \bibinfo {author} {\bibfnamefont {D.}~\bibnamefont
  {Peterseim}},\ }\bibfield  {title} {\bibinfo {title} {Nonlinear quantum
  computing by amplified encodings},\ }\href
  {https://doi.org/10.48550/arXiv.2411.16435} {\bibfield  {journal} {\bibinfo
  {journal} {arXiv preprint arXiv:2411.16435}\ } (\bibinfo {year}
  {2024})}\BibitemShut {NoStop}%
\bibitem [{\citenamefont {Byrnes}\ \emph {et~al.}(2015)\citenamefont {Byrnes},
  \citenamefont {Rosseau}, \citenamefont {Khosla}, \citenamefont {Pyrkov},
  \citenamefont {Thomasen}, \citenamefont {Mukai}, \citenamefont {Koyama},
  \citenamefont {Abdelrahman},\ and\ \citenamefont {Ilo-Okeke}}]{Byrnes2015}%
  \BibitemOpen
  \bibfield  {author} {\bibinfo {author} {\bibfnamefont {T.}~\bibnamefont
  {Byrnes}}, \bibinfo {author} {\bibfnamefont {D.}~\bibnamefont {Rosseau}},
  \bibinfo {author} {\bibfnamefont {M.}~\bibnamefont {Khosla}}, \bibinfo
  {author} {\bibfnamefont {A.}~\bibnamefont {Pyrkov}}, \bibinfo {author}
  {\bibfnamefont {A.}~\bibnamefont {Thomasen}}, \bibinfo {author}
  {\bibfnamefont {T.}~\bibnamefont {Mukai}}, \bibinfo {author} {\bibfnamefont
  {S.}~\bibnamefont {Koyama}}, \bibinfo {author} {\bibfnamefont
  {A.}~\bibnamefont {Abdelrahman}},\ and\ \bibinfo {author} {\bibfnamefont
  {E.}~\bibnamefont {Ilo-Okeke}},\ }\bibfield  {title} {\bibinfo {title}
  {Macroscopic quantum information processing using spin coherent states},\
  }\href {https://doi.org/https://doi.org/10.1016/j.optcom.2014.08.017}
  {\bibfield  {journal} {\bibinfo  {journal} {Opt. Commun.}\ }\textbf {\bibinfo
  {volume} {337}},\ \bibinfo {pages} {102} (\bibinfo {year}
  {2015})}\BibitemShut {NoStop}%
\bibitem [{\citenamefont {Xu}\ \emph {et~al.}(2022)\citenamefont {Xu},
  \citenamefont {Schmiedmayer},\ and\ \citenamefont {Sanders}}]{Xu2022PRR}%
  \BibitemOpen
  \bibfield  {author} {\bibinfo {author} {\bibfnamefont {S.}~\bibnamefont
  {Xu}}, \bibinfo {author} {\bibfnamefont {J.}~\bibnamefont {Schmiedmayer}},\
  and\ \bibinfo {author} {\bibfnamefont {B.~C.}\ \bibnamefont {Sanders}},\
  }\bibfield  {title} {\bibinfo {title} {Nonlinear quantum gates for a
  bose-einstein condensate},\ }\href
  {https://doi.org/10.1103/PhysRevResearch.4.023071} {\bibfield  {journal}
  {\bibinfo  {journal} {Phys. Rev. Res.}\ }\textbf {\bibinfo {volume} {4}},\
  \bibinfo {pages} {023071} (\bibinfo {year} {2022})}\BibitemShut {NoStop}%
\bibitem [{\citenamefont {Geller}(2024)}]{Geller2024AQT}%
  \BibitemOpen
  \bibfield  {author} {\bibinfo {author} {\bibfnamefont {M.~R.}\ \bibnamefont
  {Geller}},\ }\bibfield  {title} {\bibinfo {title} {Protocol for nonlinear
  state discrimination in rotating condensate},\ }\href
  {https://doi.org/10.1002/qute.202300431} {\bibfield  {journal} {\bibinfo
  {journal} {Adv. Quantum Technol.}\ }\textbf {\bibinfo {volume} {7}},\
  \bibinfo {pages} {2300431} (\bibinfo {year} {2024})}\BibitemShut {NoStop}%
\bibitem [{\citenamefont {Gro{\ss}ardt}(2024)}]{Grossardt2024arXiv}%
  \BibitemOpen
  \bibfield  {author} {\bibinfo {author} {\bibfnamefont {A.}~\bibnamefont
  {Gro{\ss}ardt}},\ }\bibfield  {title} {\bibinfo {title} {Nonlinear-ancilla
  aided quantum algorithm for nonlinear schr\"odinger equations},\ }\href
  {https://doi.org/10.48550/arXiv.2403.10102} {\bibfield  {journal} {\bibinfo
  {journal} {arXiv preprint arXiv:2403.10102}\ } (\bibinfo {year}
  {2024})}\BibitemShut {NoStop}%
\bibitem [{\citenamefont {Geller}(2023{\natexlab{c}})}]{Geller2023SR}%
  \BibitemOpen
  \bibfield  {author} {\bibinfo {author} {\bibfnamefont {M.~R.}\ \bibnamefont
  {Geller}},\ }\bibfield  {title} {\bibinfo {title} {Proposal for a lorenz
  qubit},\ }\href {https://doi.org/10.1038/s41598-023-40893-0} {\bibfield
  {journal} {\bibinfo  {journal} {Sci. Rep.}\ }\textbf {\bibinfo {volume}
  {13}},\ \bibinfo {pages} {14106} (\bibinfo {year}
  {2023}{\natexlab{c}})}\BibitemShut {NoStop}%
\bibitem [{\citenamefont {K\'alm\'an}\ and\ \citenamefont
  {Kiss}(2018)}]{Kalman2018PRA}%
  \BibitemOpen
  \bibfield  {author} {\bibinfo {author} {\bibfnamefont {O.}~\bibnamefont
  {K\'alm\'an}}\ and\ \bibinfo {author} {\bibfnamefont {T.}~\bibnamefont
  {Kiss}},\ }\bibfield  {title} {\bibinfo {title} {Quantum state matching of
  qubits via measurement-induced nonlinear transformations},\ }\href
  {https://doi.org/10.1103/PhysRevA.97.032125} {\bibfield  {journal} {\bibinfo
  {journal} {Phys. Rev. A}\ }\textbf {\bibinfo {volume} {97}},\ \bibinfo
  {pages} {032125} (\bibinfo {year} {2018})}\BibitemShut {NoStop}%
\bibitem [{\citenamefont {Sakaguchi}\ \emph {et~al.}(2023)\citenamefont
  {Sakaguchi}, \citenamefont {Konno}, \citenamefont {Hanamura}, \citenamefont
  {Asavanant}, \citenamefont {Takase}, \citenamefont {Ogawa}, \citenamefont
  {Marek}, \citenamefont {Filip}, \citenamefont {Yoshikawa}, \citenamefont
  {Huntington}, \citenamefont {Yonezawa},\ and\ \citenamefont
  {Furusawa}}]{Sakaguchi2023NC}%
  \BibitemOpen
  \bibfield  {author} {\bibinfo {author} {\bibfnamefont {A.}~\bibnamefont
  {Sakaguchi}}, \bibinfo {author} {\bibfnamefont {S.}~\bibnamefont {Konno}},
  \bibinfo {author} {\bibfnamefont {F.}~\bibnamefont {Hanamura}}, \bibinfo
  {author} {\bibfnamefont {W.}~\bibnamefont {Asavanant}}, \bibinfo {author}
  {\bibfnamefont {K.}~\bibnamefont {Takase}}, \bibinfo {author} {\bibfnamefont
  {H.}~\bibnamefont {Ogawa}}, \bibinfo {author} {\bibfnamefont
  {P.}~\bibnamefont {Marek}}, \bibinfo {author} {\bibfnamefont
  {R.}~\bibnamefont {Filip}}, \bibinfo {author} {\bibfnamefont {J.-i.}\
  \bibnamefont {Yoshikawa}}, \bibinfo {author} {\bibfnamefont {E.}~\bibnamefont
  {Huntington}}, \bibinfo {author} {\bibfnamefont {H.}~\bibnamefont
  {Yonezawa}},\ and\ \bibinfo {author} {\bibfnamefont {A.}~\bibnamefont
  {Furusawa}},\ }\bibfield  {title} {\bibinfo {title} {Nonlinear feedforward
  enabling quantum computation},\ }\href
  {https://doi.org/10.1038/s41467-023-39195-w} {\bibfield  {journal} {\bibinfo
  {journal} {Nat. Commun.}\ }\textbf {\bibinfo {volume} {14}},\ \bibinfo
  {pages} {3817} (\bibinfo {year} {2023})}\BibitemShut {NoStop}%
\bibitem [{\citenamefont {Yang}\ \emph {et~al.}(2008)\citenamefont {Yang},
  \citenamefont {Wen-Hong}, \citenamefont {Cun-Lin},\ and\ \citenamefont
  {Gui-Lu}}]{Yang2008CTP}%
  \BibitemOpen
  \bibfield  {author} {\bibinfo {author} {\bibfnamefont {L.}~\bibnamefont
  {Yang}}, \bibinfo {author} {\bibfnamefont {Z.}~\bibnamefont {Wen-Hong}},
  \bibinfo {author} {\bibfnamefont {Z.}~\bibnamefont {Cun-Lin}},\ and\ \bibinfo
  {author} {\bibfnamefont {L.}~\bibnamefont {Gui-Lu}},\ }\bibfield  {title}
  {\bibinfo {title} {Quantum computation with nonlinear optics},\ }\href
  {https://doi.org/10.1088/0253-6102/49/1/23} {\bibfield  {journal} {\bibinfo
  {journal} {Commun. Theor. Phys.}\ }\textbf {\bibinfo {volume} {49}},\
  \bibinfo {pages} {107} (\bibinfo {year} {2008})}\BibitemShut {NoStop}%
\bibitem [{\citenamefont {Chang}\ \emph {et~al.}(2014)\citenamefont {Chang},
  \citenamefont {Vuleti{\'{c}}},\ and\ \citenamefont {Lukin}}]{Chang2014}%
  \BibitemOpen
  \bibfield  {author} {\bibinfo {author} {\bibfnamefont {D.~E.}\ \bibnamefont
  {Chang}}, \bibinfo {author} {\bibfnamefont {V.}~\bibnamefont
  {Vuleti{\'{c}}}},\ and\ \bibinfo {author} {\bibfnamefont {M.~D.}\
  \bibnamefont {Lukin}},\ }\bibfield  {title} {\bibinfo {title} {{Quantum
  nonlinear optics — photon by photon}},\ }\href
  {https://doi.org/10.1038/nphoton.2014.192} {\bibfield  {journal} {\bibinfo
  {journal} {Nat. Photon.}\ }\textbf {\bibinfo {volume} {8}},\ \bibinfo {pages}
  {685} (\bibinfo {year} {2014})}\BibitemShut {NoStop}%
\bibitem [{\citenamefont {Gu}\ \emph {et~al.}(2016)\citenamefont {Gu},
  \citenamefont {Zhao}, \citenamefont {Baev}, \citenamefont {Yong},
  \citenamefont {Wen},\ and\ \citenamefont {Prasad}}]{Gu2016AOP}%
  \BibitemOpen
  \bibfield  {author} {\bibinfo {author} {\bibfnamefont {B.}~\bibnamefont
  {Gu}}, \bibinfo {author} {\bibfnamefont {C.}~\bibnamefont {Zhao}}, \bibinfo
  {author} {\bibfnamefont {A.}~\bibnamefont {Baev}}, \bibinfo {author}
  {\bibfnamefont {K.-T.}\ \bibnamefont {Yong}}, \bibinfo {author}
  {\bibfnamefont {S.}~\bibnamefont {Wen}},\ and\ \bibinfo {author}
  {\bibfnamefont {P.~N.}\ \bibnamefont {Prasad}},\ }\bibfield  {title}
  {\bibinfo {title} {Molecular nonlinear optics: recent advances and
  applications},\ }\href {https://doi.org/10.1364/AOP.8.000328} {\bibfield
  {journal} {\bibinfo  {journal} {Adv. Opt. Photon.}\ }\textbf {\bibinfo
  {volume} {8}},\ \bibinfo {pages} {328} (\bibinfo {year} {2016})}\BibitemShut
  {NoStop}%
\bibitem [{\citenamefont {Scala}\ \emph {et~al.}(2024)\citenamefont {Scala},
  \citenamefont {Nigro},\ and\ \citenamefont {Gerace}}]{Scala2024CP}%
  \BibitemOpen
  \bibfield  {author} {\bibinfo {author} {\bibfnamefont {F.}~\bibnamefont
  {Scala}}, \bibinfo {author} {\bibfnamefont {D.}~\bibnamefont {Nigro}},\ and\
  \bibinfo {author} {\bibfnamefont {D.}~\bibnamefont {Gerace}},\ }\bibfield
  {title} {\bibinfo {title} {Deterministic entangling gates with nonlinear
  quantum photonic interferometers},\ }\href
  {https://doi.org/10.1038/s42005-024-01610-z} {\bibfield  {journal} {\bibinfo
  {journal} {Commun. Phys.}\ }\textbf {\bibinfo {volume} {7}},\ \bibinfo
  {pages} {118} (\bibinfo {year} {2024})}\BibitemShut {NoStop}%
\bibitem [{\citenamefont {del Campo}(2021)}]{Campo2021PRL}%
  \BibitemOpen
  \bibfield  {author} {\bibinfo {author} {\bibfnamefont {A.}~\bibnamefont {del
  Campo}},\ }\bibfield  {title} {\bibinfo {title} {Probing quantum speed limits
  with ultracold gases},\ }\href
  {https://doi.org/10.1103/PhysRevLett.126.180603} {\bibfield  {journal}
  {\bibinfo  {journal} {Phys. Rev. Lett.}\ }\textbf {\bibinfo {volume} {126}},\
  \bibinfo {pages} {180603} (\bibinfo {year} {2021})}\BibitemShut {NoStop}%
\bibitem [{\citenamefont {Deffner}(2022)}]{Deffner2022EPL}%
  \BibitemOpen
  \bibfield  {author} {\bibinfo {author} {\bibfnamefont {S.}~\bibnamefont
  {Deffner}},\ }\bibfield  {title} {\bibinfo {title} {Nonlinear speed-ups in
  ultracold quantum gases},\ }\href {https://doi.org/10.1209/0295-5075/ac9fed}
  {\bibfield  {journal} {\bibinfo  {journal} {EPL (Europhys. Lett.)}\ }\textbf
  {\bibinfo {volume} {140}},\ \bibinfo {pages} {48001} (\bibinfo {year}
  {2022})}\BibitemShut {NoStop}%
\bibitem [{\citenamefont {Beau}\ and\ \citenamefont {del
  Campo}(2017)}]{Beau2017PRL}%
  \BibitemOpen
  \bibfield  {author} {\bibinfo {author} {\bibfnamefont {M.}~\bibnamefont
  {Beau}}\ and\ \bibinfo {author} {\bibfnamefont {A.}~\bibnamefont {del
  Campo}},\ }\bibfield  {title} {\bibinfo {title} {Nonlinear quantum metrology
  of many-body open systems},\ }\href
  {https://doi.org/10.1103/PhysRevLett.119.010403} {\bibfield  {journal}
  {\bibinfo  {journal} {Phys. Rev. Lett.}\ }\textbf {\bibinfo {volume} {119}},\
  \bibinfo {pages} {010403} (\bibinfo {year} {2017})}\BibitemShut {NoStop}%
\bibitem [{\citenamefont {Deffner}(2025)}]{Deffner2025QST}%
  \BibitemOpen
  \bibfield  {author} {\bibinfo {author} {\bibfnamefont {S.}~\bibnamefont
  {Deffner}},\ }\bibfield  {title} {\bibinfo {title} {Towards enhanced
  precision in thermometry with nonlinear qubits},\ }\href
  {https://doi.org/10.1088/2058-9565/adac05} {\bibfield  {journal} {\bibinfo
  {journal} {Quantum Sci. Technol.}\ }\textbf {\bibinfo {volume} {10}},\
  \bibinfo {pages} {025009} (\bibinfo {year} {2025})}\BibitemShut {NoStop}%
\bibitem [{\citenamefont {Deffner}\ \emph {et~al.}(2014)\citenamefont
  {Deffner}, \citenamefont {Jarzynski},\ and\ \citenamefont {del
  Campo}}]{Deffner2014PRX}%
  \BibitemOpen
  \bibfield  {author} {\bibinfo {author} {\bibfnamefont {S.}~\bibnamefont
  {Deffner}}, \bibinfo {author} {\bibfnamefont {C.}~\bibnamefont {Jarzynski}},\
  and\ \bibinfo {author} {\bibfnamefont {A.}~\bibnamefont {del Campo}},\
  }\bibfield  {title} {\bibinfo {title} {Classical and quantum shortcuts to
  adiabaticity for scale-invariant driving},\ }\href
  {https://doi.org/10.1103/PhysRevX.4.021013} {\bibfield  {journal} {\bibinfo
  {journal} {Phys. Rev. X}\ }\textbf {\bibinfo {volume} {4}},\ \bibinfo {pages}
  {021013} (\bibinfo {year} {2014})}\BibitemShut {NoStop}%
\bibitem [{\citenamefont {Chen}\ \emph {et~al.}(2016)\citenamefont {Chen},
  \citenamefont {Ban},\ and\ \citenamefont {Hegerfeldt}}]{Chen2016PRA}%
  \BibitemOpen
  \bibfield  {author} {\bibinfo {author} {\bibfnamefont {X.}~\bibnamefont
  {Chen}}, \bibinfo {author} {\bibfnamefont {Y.}~\bibnamefont {Ban}},\ and\
  \bibinfo {author} {\bibfnamefont {G.~C.}\ \bibnamefont {Hegerfeldt}},\
  }\bibfield  {title} {\bibinfo {title} {Time-optimal quantum control of
  nonlinear two-level systems},\ }\href
  {https://doi.org/10.1103/PhysRevA.94.023624} {\bibfield  {journal} {\bibinfo
  {journal} {Phys. Rev. A}\ }\textbf {\bibinfo {volume} {94}},\ \bibinfo
  {pages} {023624} (\bibinfo {year} {2016})}\BibitemShut {NoStop}%
\bibitem [{\citenamefont {Zhu}\ \emph {et~al.}(2020)\citenamefont {Zhu},
  \citenamefont {Chen}, \citenamefont {Jauslin},\ and\ \citenamefont
  {Gu\'erin}}]{Zhu2020PRA}%
  \BibitemOpen
  \bibfield  {author} {\bibinfo {author} {\bibfnamefont {J.-J.}\ \bibnamefont
  {Zhu}}, \bibinfo {author} {\bibfnamefont {X.}~\bibnamefont {Chen}}, \bibinfo
  {author} {\bibfnamefont {H.-R.}\ \bibnamefont {Jauslin}},\ and\ \bibinfo
  {author} {\bibfnamefont {S.}~\bibnamefont {Gu\'erin}},\ }\bibfield  {title}
  {\bibinfo {title} {Robust control of unstable nonlinear quantum systems},\
  }\href {https://doi.org/10.1103/PhysRevA.102.052203} {\bibfield  {journal}
  {\bibinfo  {journal} {Phys. Rev. A}\ }\textbf {\bibinfo {volume} {102}},\
  \bibinfo {pages} {052203} (\bibinfo {year} {2020})}\BibitemShut {NoStop}%
\bibitem [{\citenamefont {Zhu}\ and\ \citenamefont {Chen}(2021)}]{Zhu2021PRA}%
  \BibitemOpen
  \bibfield  {author} {\bibinfo {author} {\bibfnamefont {J.-J.}\ \bibnamefont
  {Zhu}}\ and\ \bibinfo {author} {\bibfnamefont {X.}~\bibnamefont {Chen}},\
  }\bibfield  {title} {\bibinfo {title} {Fast-forward scaling of atom-molecule
  conversion in bose-einstein condensates},\ }\href
  {https://doi.org/10.1103/PhysRevA.103.023307} {\bibfield  {journal} {\bibinfo
   {journal} {Phys. Rev. A}\ }\textbf {\bibinfo {volume} {103}},\ \bibinfo
  {pages} {023307} (\bibinfo {year} {2021})}\BibitemShut {NoStop}%
\bibitem [{\citenamefont {Zhu}\ \emph {et~al.}(2023)\citenamefont {Zhu},
  \citenamefont {Liu}, \citenamefont {Chen},\ and\ \citenamefont
  {Gu\'erin}}]{Zhu2023PRA}%
  \BibitemOpen
  \bibfield  {author} {\bibinfo {author} {\bibfnamefont {J.-j.}\ \bibnamefont
  {Zhu}}, \bibinfo {author} {\bibfnamefont {K.}~\bibnamefont {Liu}}, \bibinfo
  {author} {\bibfnamefont {X.}~\bibnamefont {Chen}},\ and\ \bibinfo {author}
  {\bibfnamefont {S.}~\bibnamefont {Gu\'erin}},\ }\bibfield  {title} {\bibinfo
  {title} {Optimal control and ultimate bounds of 1:2 nonlinear quantum
  systems},\ }\href {https://doi.org/10.1103/PhysRevA.108.042610} {\bibfield
  {journal} {\bibinfo  {journal} {Phys. Rev. A}\ }\textbf {\bibinfo {volume}
  {108}},\ \bibinfo {pages} {042610} (\bibinfo {year} {2023})}\BibitemShut
  {NoStop}%
\bibitem [{\citenamefont {Deffner}\ and\ \citenamefont
  {Campbell}(2025)}]{Deffner2025EPL}%
  \BibitemOpen
  \bibfield  {author} {\bibinfo {author} {\bibfnamefont {S.}~\bibnamefont
  {Deffner}}\ and\ \bibinfo {author} {\bibfnamefont {S.}~\bibnamefont
  {Campbell}},\ }\bibfield  {title} {\bibinfo {title} {Suppressing excitations
  in the nonlinear landau-zener model},\ }\href
  {https://doi.org/10.1209/0295-5075/adfd7a} {\bibfield  {journal} {\bibinfo
  {journal} {EPL (Europhys. Lett.)}\ }\textbf {\bibinfo {volume} {151}},\
  \bibinfo {pages} {58001} (\bibinfo {year} {2025})}\BibitemShut {NoStop}%
\bibitem [{Note1()}]{Note1}%
  \BibitemOpen
  \bibinfo {note} {Note that Eq.~\protect \eqref {eq:var} implies $\beta
  =\partial S/\partial E$, which is identical to the thermodynamic definition
  of temperature \cite {Callen1985}.}\BibitemShut {Stop}%
\bibitem [{\citenamefont {Callen}(1985)}]{Callen1985}%
  \BibitemOpen
  \bibfield  {author} {\bibinfo {author} {\bibfnamefont {H.~B.}\ \bibnamefont
  {Callen}},\ }\href
  {https://en.wikipedia.org/wiki/Thermodynamics_and_an_Introduction_to_Thermostatistics}
  {\emph {\bibinfo {title} {Thermodynamics and an introduction to
  thermostatistics}}}\ (\bibinfo  {publisher} {Wiley},\ \bibinfo {address} {New
  York, USA},\ \bibinfo {year} {1985})\BibitemShut {NoStop}%
\bibitem [{\citenamefont {Deffner}\ and\ \citenamefont
  {Campbell}(2019)}]{Deffner2019book}%
  \BibitemOpen
  \bibfield  {author} {\bibinfo {author} {\bibfnamefont {S.}~\bibnamefont
  {Deffner}}\ and\ \bibinfo {author} {\bibfnamefont {S.}~\bibnamefont
  {Campbell}},\ }\href {https://doi.org/10.1088/2053-2571/ab21c6} {\emph
  {\bibinfo {title} {Quantum Thermodynamics}}}\ (\bibinfo  {publisher} {Morgan
  \& Claypool Publishers},\ \bibinfo {year} {2019})\BibitemShut {NoStop}%
\bibitem [{\citenamefont {Campbell}\ \emph {et~al.}(2025)\citenamefont
  {Campbell}, \citenamefont {D'Amico}, \citenamefont {Ciampini} \emph
  {et~al.}}]{campbell2025roadmap}%
  \BibitemOpen
  \bibfield  {author} {\bibinfo {author} {\bibfnamefont {S.}~\bibnamefont
  {Campbell}}, \bibinfo {author} {\bibfnamefont {I.}~\bibnamefont {D'Amico}},
  \bibinfo {author} {\bibfnamefont {M.~A.}\ \bibnamefont {Ciampini}}, \emph
  {et~al.},\ }\bibfield  {title} {\bibinfo {title} {Roadmap on quantum
  thermodynamics},\ }\href {https://doi.org/0.48550/arXiv.2504.20145}
  {\bibfield  {journal} {\bibinfo  {journal} {arXiv preprint arXiv:2504.20145}\
  } (\bibinfo {year} {2025})}\BibitemShut {NoStop}%
\bibitem [{\citenamefont {Quan}\ \emph {et~al.}(2007)\citenamefont {Quan},
  \citenamefont {Liu}, \citenamefont {Sun},\ and\ \citenamefont
  {Nori}}]{Quan2007PRE}%
  \BibitemOpen
  \bibfield  {author} {\bibinfo {author} {\bibfnamefont {H.~T.}\ \bibnamefont
  {Quan}}, \bibinfo {author} {\bibfnamefont {Y.-x.}\ \bibnamefont {Liu}},
  \bibinfo {author} {\bibfnamefont {C.~P.}\ \bibnamefont {Sun}},\ and\ \bibinfo
  {author} {\bibfnamefont {F.}~\bibnamefont {Nori}},\ }\bibfield  {title}
  {\bibinfo {title} {Quantum thermodynamic cycles and quantum heat engines},\
  }\href {https://doi.org/10.1103/PhysRevE.76.031105} {\bibfield  {journal}
  {\bibinfo  {journal} {Phys. Rev. E}\ }\textbf {\bibinfo {volume} {76}},\
  \bibinfo {pages} {031105} (\bibinfo {year} {2007})}\BibitemShut {NoStop}%
\bibitem [{\citenamefont {Smith}\ \emph {et~al.}(2020)\citenamefont {Smith},
  \citenamefont {Pal},\ and\ \citenamefont {Deffner}}]{Smith2020JNET}%
  \BibitemOpen
  \bibfield  {author} {\bibinfo {author} {\bibfnamefont {Z.}~\bibnamefont
  {Smith}}, \bibinfo {author} {\bibfnamefont {P.~S.}\ \bibnamefont {Pal}},\
  and\ \bibinfo {author} {\bibfnamefont {S.}~\bibnamefont {Deffner}},\
  }\bibfield  {title} {\bibinfo {title} {Endoreversible otto engines at maximal
  power},\ }\href {https://doi.org/doi:10.1515/jnet-2020-0039} {\bibfield
  {journal} {\bibinfo  {journal} {J. Non-Equilib. Thermodyn.}\ }\textbf
  {\bibinfo {volume} {45}},\ \bibinfo {pages} {305} (\bibinfo {year}
  {2020})}\BibitemShut {NoStop}%
\bibitem [{\citenamefont {Abah}\ \emph {et~al.}(2012)\citenamefont {Abah},
  \citenamefont {Ro\ss{}nagel}, \citenamefont {Jacob}, \citenamefont {Deffner},
  \citenamefont {Schmidt-Kaler}, \citenamefont {Singer},\ and\ \citenamefont
  {Lutz}}]{Abah2012PRL}%
  \BibitemOpen
  \bibfield  {author} {\bibinfo {author} {\bibfnamefont {O.}~\bibnamefont
  {Abah}}, \bibinfo {author} {\bibfnamefont {J.}~\bibnamefont {Ro\ss{}nagel}},
  \bibinfo {author} {\bibfnamefont {G.}~\bibnamefont {Jacob}}, \bibinfo
  {author} {\bibfnamefont {S.}~\bibnamefont {Deffner}}, \bibinfo {author}
  {\bibfnamefont {F.}~\bibnamefont {Schmidt-Kaler}}, \bibinfo {author}
  {\bibfnamefont {K.}~\bibnamefont {Singer}},\ and\ \bibinfo {author}
  {\bibfnamefont {E.}~\bibnamefont {Lutz}},\ }\bibfield  {title} {\bibinfo
  {title} {Single-ion heat engine at maximum power},\ }\href
  {https://doi.org/10.1103/PhysRevLett.109.203006} {\bibfield  {journal}
  {\bibinfo  {journal} {Phys. Rev. Lett.}\ }\textbf {\bibinfo {volume} {109}},\
  \bibinfo {pages} {203006} (\bibinfo {year} {2012})}\BibitemShut {NoStop}%
\bibitem [{\citenamefont {Deffner}(2018)}]{Deffner2018Entropy}%
  \BibitemOpen
  \bibfield  {author} {\bibinfo {author} {\bibfnamefont {S.}~\bibnamefont
  {Deffner}},\ }\bibfield  {title} {\bibinfo {title} {Efficiency of harmonic
  quantum otto engines at maximal power},\ }\href
  {https://doi.org/10.3390/e20110875} {\bibfield  {journal} {\bibinfo
  {journal} {Entropy}\ }\textbf {\bibinfo {volume} {20}},\ \bibinfo {pages}
  {875} (\bibinfo {year} {2018})}\BibitemShut {NoStop}%
\bibitem [{\citenamefont {Myers}\ and\ \citenamefont
  {Deffner}(2020)}]{Myers2020PRE}%
  \BibitemOpen
  \bibfield  {author} {\bibinfo {author} {\bibfnamefont {N.~M.}\ \bibnamefont
  {Myers}}\ and\ \bibinfo {author} {\bibfnamefont {S.}~\bibnamefont
  {Deffner}},\ }\bibfield  {title} {\bibinfo {title} {Bosons outperform
  fermions: The thermodynamic advantage of symmetry},\ }\href
  {https://doi.org/10.1103/PhysRevE.101.012110} {\bibfield  {journal} {\bibinfo
   {journal} {Phys. Rev. E}\ }\textbf {\bibinfo {volume} {101}},\ \bibinfo
  {pages} {012110} (\bibinfo {year} {2020})}\BibitemShut {NoStop}%
\bibitem [{\citenamefont {Myers}\ and\ \citenamefont
  {Deffner}(2021)}]{Myers2021PRXQ}%
  \BibitemOpen
  \bibfield  {author} {\bibinfo {author} {\bibfnamefont {N.~M.}\ \bibnamefont
  {Myers}}\ and\ \bibinfo {author} {\bibfnamefont {S.}~\bibnamefont
  {Deffner}},\ }\bibfield  {title} {\bibinfo {title} {Thermodynamics of
  statistical anyons},\ }\href {https://doi.org/10.1103/PRXQuantum.2.040312}
  {\bibfield  {journal} {\bibinfo  {journal} {PRX Quantum}\ }\textbf {\bibinfo
  {volume} {2}},\ \bibinfo {pages} {040312} (\bibinfo {year}
  {2021})}\BibitemShut {NoStop}%
\bibitem [{\citenamefont {Myers}\ \emph
  {et~al.}(2021{\natexlab{a}})\citenamefont {Myers}, \citenamefont {Abah},\
  and\ \citenamefont {Deffner}}]{Myers2021NJP}%
  \BibitemOpen
  \bibfield  {author} {\bibinfo {author} {\bibfnamefont {N.~M.}\ \bibnamefont
  {Myers}}, \bibinfo {author} {\bibfnamefont {O.}~\bibnamefont {Abah}},\ and\
  \bibinfo {author} {\bibfnamefont {S.}~\bibnamefont {Deffner}},\ }\bibfield
  {title} {\bibinfo {title} {Quantum otto engines at relativistic energies},\
  }\href {https://doi.org/10.1088/1367-2630/ac2756} {\bibfield  {journal}
  {\bibinfo  {journal} {New J. Phys.}\ }\textbf {\bibinfo {volume} {23}},\
  \bibinfo {pages} {105001} (\bibinfo {year} {2021}{\natexlab{a}})}\BibitemShut
  {NoStop}%
\bibitem [{\citenamefont {Hoffmann}\ \emph {et~al.}(1997)\citenamefont
  {Hoffmann}, \citenamefont {Burzler},\ and\ \citenamefont
  {Schubert}}]{Hoffmann1997}%
  \BibitemOpen
  \bibfield  {author} {\bibinfo {author} {\bibfnamefont {K.~H.}\ \bibnamefont
  {Hoffmann}}, \bibinfo {author} {\bibfnamefont {J.~M.}\ \bibnamefont
  {Burzler}},\ and\ \bibinfo {author} {\bibfnamefont {S.}~\bibnamefont
  {Schubert}},\ }\bibfield  {title} {\bibinfo {title} {Endoreversible
  thermodynamics},\ }\href {https://doi.org/10.1515/jnet.1997.22.4.311}
  {\bibfield  {journal} {\bibinfo  {journal} {J. Non-Equilib. Thermodyn.}\
  }\textbf {\bibinfo {volume} {22}},\ \bibinfo {pages} {311} (\bibinfo {year}
  {1997})}\BibitemShut {NoStop}%
\bibitem [{\citenamefont {Curzon}\ and\ \citenamefont
  {Ahlborn}(1975)}]{Curzon1975AJP}%
  \BibitemOpen
  \bibfield  {author} {\bibinfo {author} {\bibfnamefont {F.~L.}\ \bibnamefont
  {Curzon}}\ and\ \bibinfo {author} {\bibfnamefont {B.}~\bibnamefont
  {Ahlborn}},\ }\bibfield  {title} {\bibinfo {title} {{Efficiency of a {Carnot}
  engine at maximum power output}},\ }\href {https://doi.org/10.1119/1.10023}
  {\bibfield  {journal} {\bibinfo  {journal} {Am. J. Phys.}\ }\textbf {\bibinfo
  {volume} {43}},\ \bibinfo {pages} {22} (\bibinfo {year} {1975})}\BibitemShut
  {NoStop}%
\bibitem [{\citenamefont {Ferketic}\ and\ \citenamefont
  {Deffner}(2023)}]{Ferketic2023EPL}%
  \BibitemOpen
  \bibfield  {author} {\bibinfo {author} {\bibfnamefont {E.~E.}\ \bibnamefont
  {Ferketic}}\ and\ \bibinfo {author} {\bibfnamefont {S.}~\bibnamefont
  {Deffner}},\ }\bibfield  {title} {\bibinfo {title} {Boosting thermodynamic
  performance by bending space-time},\ }\href
  {https://doi.org/10.1209/0295-5075/acad9c} {\bibfield  {journal} {\bibinfo
  {journal} {EPL (Europhys. Lett.)}\ }\textbf {\bibinfo {volume} {141}},\
  \bibinfo {pages} {19001} (\bibinfo {year} {2023})}\BibitemShut {NoStop}%
\bibitem [{\citenamefont {Myers}\ \emph
  {et~al.}(2021{\natexlab{b}})\citenamefont {Myers}, \citenamefont {McCready},\
  and\ \citenamefont {Deffner}}]{Myers2021Symmetry}%
  \BibitemOpen
  \bibfield  {author} {\bibinfo {author} {\bibfnamefont {N.~M.}\ \bibnamefont
  {Myers}}, \bibinfo {author} {\bibfnamefont {J.}~\bibnamefont {McCready}},\
  and\ \bibinfo {author} {\bibfnamefont {S.}~\bibnamefont {Deffner}},\
  }\bibfield  {title} {\bibinfo {title} {Quantum heat engines with singular
  interactions},\ }\href {https://doi.org/10.3390/sym13060978} {\bibfield
  {journal} {\bibinfo  {journal} {Symmetry}\ }\textbf {\bibinfo {volume}
  {13}},\ \bibinfo {pages} {978} (\bibinfo {year}
  {2021}{\natexlab{b}})}\BibitemShut {NoStop}%
\bibitem [{\citenamefont {Myers}\ \emph {et~al.}(2022)\citenamefont {Myers},
  \citenamefont {Peña}, \citenamefont {Negrete}, \citenamefont {Vargas},
  \citenamefont {De~Chiara},\ and\ \citenamefont {Deffner}}]{Myers2022NJP}%
  \BibitemOpen
  \bibfield  {author} {\bibinfo {author} {\bibfnamefont {N.~M.}\ \bibnamefont
  {Myers}}, \bibinfo {author} {\bibfnamefont {F.~J.}\ \bibnamefont {Peña}},
  \bibinfo {author} {\bibfnamefont {O.}~\bibnamefont {Negrete}}, \bibinfo
  {author} {\bibfnamefont {P.}~\bibnamefont {Vargas}}, \bibinfo {author}
  {\bibfnamefont {G.}~\bibnamefont {De~Chiara}},\ and\ \bibinfo {author}
  {\bibfnamefont {S.}~\bibnamefont {Deffner}},\ }\bibfield  {title} {\bibinfo
  {title} {Boosting engine performance with bose–einstein condensation},\
  }\href {https://doi.org/10.1088/1367-2630/ac47cc} {\bibfield  {journal}
  {\bibinfo  {journal} {New J. Phys.}\ }\textbf {\bibinfo {volume} {24}},\
  \bibinfo {pages} {025001} (\bibinfo {year} {2022})}\BibitemShut {NoStop}%
\bibitem [{\citenamefont {Myers}\ \emph {et~al.}(2023)\citenamefont {Myers},
  \citenamefont {Peña}, \citenamefont {Cortés},\ and\ \citenamefont
  {Vargas}}]{Myers2023Nanomat}%
  \BibitemOpen
  \bibfield  {author} {\bibinfo {author} {\bibfnamefont {N.~M.}\ \bibnamefont
  {Myers}}, \bibinfo {author} {\bibfnamefont {F.~J.}\ \bibnamefont {Peña}},
  \bibinfo {author} {\bibfnamefont {N.}~\bibnamefont {Cortés}},\ and\ \bibinfo
  {author} {\bibfnamefont {P.}~\bibnamefont {Vargas}},\ }\bibfield  {title}
  {\bibinfo {title} {Multilayer graphene as an endoreversible otto engine},\
  }\href {https://doi.org/10.3390/nano13091548} {\bibfield  {journal} {\bibinfo
   {journal} {Nanomaterials}\ }\textbf {\bibinfo {volume} {13}},\ \bibinfo
  {pages} {1548} (\bibinfo {year} {2023})}\BibitemShut {NoStop}%
\bibitem [{\citenamefont {Peña}\ \emph {et~al.}(2023)\citenamefont {Peña},
  \citenamefont {Myers}, \citenamefont {Órdenes}, \citenamefont
  {Albarrán-Arriagada},\ and\ \citenamefont {Vargas}}]{Pena2023Entropy}%
  \BibitemOpen
  \bibfield  {author} {\bibinfo {author} {\bibfnamefont {F.~J.}\ \bibnamefont
  {Peña}}, \bibinfo {author} {\bibfnamefont {N.~M.}\ \bibnamefont {Myers}},
  \bibinfo {author} {\bibfnamefont {D.}~\bibnamefont {Órdenes}}, \bibinfo
  {author} {\bibfnamefont {F.}~\bibnamefont {Albarrán-Arriagada}},\ and\
  \bibinfo {author} {\bibfnamefont {P.}~\bibnamefont {Vargas}},\ }\bibfield
  {title} {\bibinfo {title} {Enhanced efficiency at maximum power in a
  fock–darwin model quantum dot engine},\ }\href
  {https://doi.org/10.3390/e25030518} {\bibfield  {journal} {\bibinfo
  {journal} {Entropy}\ }\textbf {\bibinfo {volume} {25}},\ \bibinfo {pages}
  {518} (\bibinfo {year} {2023})}\BibitemShut {NoStop}%
\bibitem [{\citenamefont {Behrendt}\ and\ \citenamefont
  {Deffner}(2025)}]{Behrendt2025entropy}%
  \BibitemOpen
  \bibfield  {author} {\bibinfo {author} {\bibfnamefont {G.}~\bibnamefont
  {Behrendt}}\ and\ \bibinfo {author} {\bibfnamefont {S.}~\bibnamefont
  {Deffner}},\ }\bibfield  {title} {\bibinfo {title} {Endoreversible stirling
  cycles: Plasma engines at maximal power},\ }\href
  {https://doi.org/10.3390/e27080807} {\bibfield  {journal} {\bibinfo
  {journal} {Entropy}\ }\textbf {\bibinfo {volume} {27}},\ \bibinfo {pages}
  {807} (\bibinfo {year} {2025})}\BibitemShut {NoStop}%
\end{thebibliography}%

\end{document}